\documentclass[aps,prl,reprint,superscriptaddress,longbibliography]{revtex4-2}
\usepackage[T1]{fontenc}
\usepackage[utf8]{inputenc}
\usepackage{amsmath,amssymb,bm,mathtools,slashed}
\usepackage{graphicx}
\usepackage{booktabs}
\usepackage{longtable}
\usepackage{microtype}
\usepackage{hyperref}
\usepackage{tikz}
\usetikzlibrary{arrows.meta, positioning, calc}
\hypersetup{colorlinks=true,citecolor=blue,linkcolor=blue,urlcolor=blue}

\newcommand{\Tr}{\mathrm{Tr}}
\newcommand{\dd}{\mathrm{d}}
\newcommand{\cL}{\mathcal{L}}
\newcommand{\cH}{\mathcal{H}}

\newcommand{\LLL}{\mathrm{LLL}}
\newcommand{\HLL}{\mathrm{HLL}}

\newcommand{\Z}{{\mathbb Z}}
\newcommand{\R}{{\mathbb R}}

\newcommand{\PH}{\mathrm{PH}}
\newcommand{\sgn}{\operatorname{sgn}}
\newcommand{\ep}{\mathrm{ep}}
\newcommand{\He}{\mathrm{He}}

\def\cA{{\mathcal A}}

\def\cH{{\mathcal H}}

\def\cL{{\mathcal L}}

\def\cO{{\mathcal O}}

\begin{document}

\title{The Effective Field Theory Origin of the Fuzzy Sphere Method}
\author{Mykola Dedushenko}
\affiliation{
	Center for Mathematics and Interdisciplinary Sciences, Fudan University, Shanghai, 200433, China
}
\affiliation{
	Shanghai Institute for Mathematics and Interdisciplinary Sciences (SIMIS), Shanghai, 200433, China
}

\begin{abstract}
Fuzzy sphere Hamiltonians for (2+1)D conformal field theories are usually written directly on the lowest Landau level. We introduce a framework organizing them as descendants of local ``mother theories'' on $S^2\times\mathbb R$. We start with the example of Ising mother theory built from two charged Dirac fermions coupled to a real mediator $\phi$. Integrating out $\phi$ and the higher Landau levels produces the fuzzy Hamiltonian, while integrating out the fermions gives a conventional local $\phi$ effective field theory flowing to the Ising point. Mild modifications of the same theory reach the Gaussian point (tricritical route) and the Lee-Yang continuation. Exploring the space of mother theories, we next promote $\phi$ to a complex field $\Phi$ and construct a new example, whose descendant is an easy-plane ferromagnet on the fuzzy sphere. Its spontaneous symmetry breaking phase gives a pseudo-Goldstone relativistic scalar, similar to the model introduced by Y.-C.~He, while its critical point is non-relativistic, providing the first example of $z=2$ scaling on the fuzzy sphere. Remarkably, it admits exact one-magnon solutions, and the $z=2$ critical point can be found analytically. We then outline a rich landscape of further applications of our approach.
\end{abstract}

\maketitle

\section{Introduction}
The fuzzy sphere \cite{Madore1992,Haldane1983,GrosseKlimcikPresnajder1996} approach to 3D CFT was initiated in \cite{ZhuHanHuffmanHofmannHe2023}. Unlike the older 2D applications, this method takes place on $S^2\times \R_{\rm t}$, where only the space $S^2$ is made non-commutative, replacing $C^\infty(S^2)$ by a finite-dimensional algebra of fixed-size square matrices, explicitly breaking the Lorentz symmetry, and approximating a 3D QFT by a matrix quantum mechanics (QM). The spatial $SO(3)$ remains preserved, which is the key advantage  over the lattice regularization, making the fuzzy $S^2$ an $SO(3)$-invariant analog of discretized $S^2$. One can write fuzzy analogs for many lattice models, with the role of sites played by fuzzy orbitals (i.e., monopole sphere harmonics). The prime example is the fuzzy Ising model \cite{ZhuHanHuffmanHofmannHe2023}, built using the isospin-$\frac12$ fermions, exhibiting the usual 3D Ising CFT at its quantum critical point. 

The new method was met with enthusiasm: powered by the Julia package FuzzifiED \cite{Zhou2025FuzzifiED}, more models were constructed~\cite{HanHuZhu2024,ZhouHuZhuHe2024,ChenEtAl2024SO5,YangEtAl2026O4,ZhouHe2025SpN,YangEtAl2025Potts,FanDongVishwanath2026,ArguelloCruzEtAl2025,EliasMiroDelouche2025,He2025FreeScalar,TaylorEtAl2026,GuoEtAl2026,DeyEtAl2026O3,DeyEtAl2026O2,Stergiou2026,ZhouGaiottoHe2025,VoineaEtAl2026Majorana,TangEtAl2026SUSY,ZhouWangHe2025,HuffmanEtAl2026}, and the technique was extended beyond spectra \cite{HuHeZhu2023,HanHuZhuHe2023,HofmannEtAl2024,HuEtAl2024Defects,ZhouEtAl2024GFunction,CuomoHeKomargodski2024,ZhouZou2025Surface,Dedushenko2024,SarmaEtAl2026,FengWang2026,YangHuChenZhu2026Defect,HuZhuHe2025F,FardelliEtAl2025,Fan2024Generators,LauchliEtAl2025,DongEtAl2025Crosscap,Wiese2026,HaoZhuQi2026,FardelliEtAl2026OPE,EckWang2026,RamirezCamascaEtAl2026,Baran2026} and in some other ways \cite{FardelliEtAl2026IFT,HanZhu2025Circle,MengZhaoZhuWu2026CP2,MengHuZhu2026S3}. A few questions, however, have been coming up: Is it possible to understand it from the effective field theory (EFT) point of view? How can we systematically construct fuzzy models? (Existing constructions often rely on physical intuition and numerical evidence, though some recent progress on systematizing the building blocks was made in \cite{ZhouGaiottoHe2025}.) Why does the method work so well? Why is the IR CFT Lorentz-invariant (while $\cH_{\rm fuzzy}$ is not): Could a more general scaling $t\mapsto \lambda^z t$, $\vec{x}\mapsto \lambda \vec{x}$ with $z\neq 1$ emerge instead? (Many-body systems with $z\neq 1$ critical points are well-known \cite{FisherWeichmanGrinsteinFisher1989,SachdevSenthilShankar1994,NikuniOshikawaOosawaTanaka2000,ArdonneFendleyFradkin2004,Hertz1976,Millis1993,OganesyanKivelsonFradkin2001,Fisher1995RandomTFIM}.)

This work answers these questions. We build a field-theoretic foundation for the fuzzy method, thus answering the question about its EFT origins; provide a way to systematically build new models; understand emergent Lorentz symmetry of the CFT, and give an example of a fuzzy model with non-relativistic ($z=2$) critical point. We also open the path to generalizing to gapped QFTs. We first explain the organizing idea of mother theory and develop it for Ising CFT, including the tri-critical free scalar point and the Lee-Yang deformation. Then we describe a new easy-plane family of fuzzy models, which includes another realization of free scalar (closely related to \cite{He2025FreeScalar}) and an example of $z=2$ criticality. We conclude by listing further directions, including those addressed in a companion paper. The End Matter and Supplemental Materials \cite{SupplementalMaterial} contain many technical details.

\section{Organizing idea}
We propose that behind each fuzzy sphere model, there is local QFT, its \emph{mother theory}, manifestly connecting the model Hamiltonian $\cH_{\rm fuzzy}$ to the local CFT. It consists of heavy charged fields $\Psi$ and light neutral mediators $\Phi$,
\begin{equation}
\cL_{\rm mother}[\Psi, \Phi].
\end{equation}
Put $\cL_{\rm mother}$ on $S^2\times \R$ with large magnetic flux through $S^2$ (under which only $\Psi$ are charged). The first qualifying property of the mother theory is that integrating out $\Psi$ produces a local EFT of $\Phi$ on $S^2\times \R$ whose IR critical point is the CFT we wish to study. The second qualifying property is that integrating out $\Phi$ and projecting on the lowest Landau level (LLL) for $\Psi$ (which \emph{really} means integrating out higher Landau levels (HLLs)), produces the desired fuzzy sphere model. Thus the key properties of mother theory are summarized in the diagram:
\begin{tikzpicture}[
	basebox/.style={
		draw,
		rounded corners,
		minimum height=0.8cm,
		inner ysep=1.5pt,
		align=center
	},
	motherbox/.style={
		basebox,
		minimum width=0pt,   
		inner xsep=5pt
	},
	lowerbox/.style={
		basebox,
		minimum width=3.0cm,
		inner xsep=6pt
	},
	arrow/.style={
		-{Latex[length=2mm]},
		thick,
		rounded corners=1.5pt
	},
	arrowlabel/.style={
		font=\scriptsize,
		inner sep=1pt,
		fill=white
	}
	]
	
	\node[motherbox] (mother) at (0.4,1.55)
	{$\cL_{\rm mother}[\Psi,\Phi]$};
	
	\node[lowerbox] (fuzzy) at (-2.35,0.3)
	{$\cH_{\rm fuzzy}$};
	
	\node[lowerbox] (eft) at (2.95,0.3)
	{$\cL_{\rm local\ EFT}
		\xrightarrow{\rm RG}
		{\rm CFT}$};
	
	\coordinate (lefttarget)
	at ($(fuzzy.north west)+(0.42cm,0)$);
	
	\coordinate (righttarget)
	at ($(eft.north east)+(-0.42cm,0)$);
	
	\coordinate (leftbend)
	at (lefttarget |- mother.south west);
	
	\coordinate (rightbend)
	at (righttarget |- mother.south east);
	
	\draw[arrow]
	(mother.south west)
	--
	node[pos=0.50, above=2pt, arrowlabel]
	{integrate out $\Phi$}
	node[pos=0.50, below=2pt, arrowlabel]
	{project to LLL}
	(leftbend)
	--
	(lefttarget);
	
	\draw[arrow]
	(mother.south east)
	--
	node[pos=0.50, above=2pt, arrowlabel]
	{integrate out $\Psi$}
	(rightbend)
	--
	(righttarget);
	
\end{tikzpicture}

To make this more precise, assume the separation of scales between the masses of $\Psi$ and the Landau gap:
\begin{tikzpicture}[
scale=1.0,
level/.style={thick},
labeltext/.style={anchor=west, align=left}
]

\draw[->, thick] (0,0) -- (0,1.4)
node[above] {energy};

\draw[level] (-0.15,0) -- (2.2,0);
\node[anchor=east] at (-0.25,0) {$0$};
\node[labeltext] at (2.7,0) {light mediators $\Phi$};

\draw[level] (-0.15,0.5) -- (2.2,0.5);
\node[anchor=east] at (-0.25,0.5) {$m$};
\node[labeltext] at (2.7,0.5) {mass scale of fields $\Psi$};

\draw[level] (-0.15,1.2) -- (2.2,1.2);
\node[anchor=east] at (-0.25,1.2) {$\Delta_{\rm LL}$};
\node[labeltext] at (2.7,1.2) {higher Landau levels};

\end{tikzpicture}\\
At energies $\ll m$, we naturally get a local EFT of light degrees of freedom (dof) $\Phi$. However, below the cutoff $\Lambda$, such that $m\ll\Lambda\ll \Delta_{\rm LL}$, the natural description retains both $\Phi$ and the LLL-projected fields $\Psi$. The contribution of HLLs is suppressed in the large $\Delta_{\rm LL}$ limit. Even though $\Phi$'s are lighter than $\Psi$, we can integrate them out, producing a \emph{non-local} theory of LLL-projected $\Psi$. This gives the fuzzy Hamiltonian $\cH_{\rm fuzzy}$, typically a finite QM model since the LLL is finite-dimensional. In such a description, the gapless CFT dof $\Phi$ becomes a collective excitation, the bilinear of gapped charged dof:
\begin{equation}
\Phi \sim \bar\Psi\Psi.
\end{equation}

The mother theory perspective makes interaction channels, symmetries, and criticality tunings transparent. It also puts the dynamical exponent $z$ under control, allowing to read it off from the EFT analysis.

\section{3D Ising mother theory}

We start with the Ising CFT. Let $\Psi=(\psi_1,\psi_2)^T$ be two charge $+1$ Dirac fermions (flavor doublet) on $S^2\times\mathbb R$ in a $U(1)$ monopole background $A_\mu$ of flux $N$, and define
\begin{equation}
 n^a=\bar\Psi\sigma_F^a\Psi,\qquad
 n^0=\bar\Psi\Psi,
\label{eq:bilinears}
\end{equation}
where $\sigma_F^a$ are Pauli matrices in flavor space. The minimal Ising mother also has a real scalar $\phi$, and the action is:
\begin{eqnarray}\label{eq:IsingMother}
\cL_I={}&\bar\Psi \gamma^\mu D_\mu[A]\Psi
+m n^x+g\phi n^z+\frac12\phi K^{-1}_\phi\phi,\cr
&K^{-1}_\phi=r_\phi-\kappa_\phi\nabla^2.
\end{eqnarray}
At $\kappa_\phi=0$, $\phi$ is an auxiliary Hubbard-Stratonovich (HS) mediator decoupling the four-Fermi interaction $n^z n^z$, called the Ising channel interaction. For nonzero bare kinetic coefficient $\kappa_\phi$,  $\phi$ is dynamical, although we continue to call it an HS field for brevity. Symmetries of the model include: (1)\nobreakdash\ $SU(2)$ spatial rotations; (2) $\Z_2$:
\begin{equation}
\text{Ising }\Z_2:\qquad \phi\mapsto-\phi,
\qquad
\Psi\mapsto\sigma_F^x\Psi,
\label{eq:IsingZ2}
\end{equation}
(3) $\widetilde{CP}$, $\widetilde{CT}$ discrete symmetries, which we now define. 
A pair of free Dirac fermions has a $U(2)_F$ flavor symmetry, charge conjugation $C$, time-reversal $T$, space parity $P$. The diagonal $U(1)_V\subset U(2)_F$ is fermion number coupled to the background $A_\mu$. The terms $m n^x + g\phi n^z$ are parity-odd in 3D, explicitly violating $P$, $T$ and $SU(2)_F$. The following combinations survive at $A_\mu=0$:
\begin{equation}
\widetilde{P} = (i\sigma^y_F)P,\quad \widetilde{T} = (i\sigma^y_F)T. 
\end{equation}
The monopole background $A_\mu$ then breaks $C$, $\widetilde{P}$ and $\widetilde{T}$ individually, only preserving the products:
\begin{equation}
\widetilde{CP} = C (i\sigma^y_F) P,\quad \widetilde{CT} = C (i\sigma^y_F) T.
\end{equation}

Let us verify that this is indeed the mother theory for Ising. The dynamical fermion mass matrix,
\begin{equation}
m\sigma^x_F + g\phi \sigma^z_F,
\end{equation}
has the eigenvalues $\pm\sqrt{m^2 + g^2\phi^2}$ that remain nonzero so long as $m\neq 0$. Thus we can integrate fermions out, producing a local scalar EFT constrained by the Ising $\Z_2$ symmetry $\phi\mapsto -\phi$:
\begin{eqnarray}
\cL_{\rm loc. EFT}&=
\frac{Z_\phi}{2}(\partial\phi)^2
+\frac{r_2}{2}\phi^2
+\frac{u_4}{4!}\phi^4
+  \cdots .
\label{eq:scalarEFT}
\end{eqnarray}
The $\widetilde{CP}$ and $\widetilde{CT}$ act as $P$ and $T$ on $\phi$. The exact induced potential in $\R^3$ is $\frac1{3\pi}\left[ (m^2+g^2\phi^2)^{3/2} - |m|^3 \right]$ (up to the mass counterterm), and the first few coefficients are:
\begin{eqnarray}\label{eq:scalarEFT-coefficients}
&Z_\phi = \kappa_\phi + \frac{g^2}{4\pi|m|},\  
r_2=r_\phi + \frac{g^2|m|}{\pi},\ u_4=\frac{3g^4}{\pi|m|}, 
\dots
\end{eqnarray}
Note that $u_4>0$, so tuning $r_2$ indeed reaches the ordinary 3D Ising Wilson--Fisher fixed point~\cite{WilsonFisher1972}.

Actually, we need this conclusion on fluxed $S^2\times \R$. A technical integration over the LLs of $\Psi$ is described in the Supplemental Material \cite{SupplementalMaterial}. It gives the same EFT \eqref{eq:scalarEFT}, with more complicated coefficients, yet still $u_4>0$, so the mother theory obeys the first qualifying property.

To obtain $\cH_{\rm fuzzy}$, we first integrate out HLLs. The gap $\Delta_{LL}\sim \sqrt{N}$ suppresses their effect, except for the renormalization of $r_\phi$, leading to the intermediate Lagrangian:
\begin{align}\label{eq:intermediate}
\cL_{int}= \bar\Psi_{\rm LLL} \gamma^\tau \partial_\tau \Psi_{\rm LLL} + m n^x_{\rm LLL} + g\phi n^z_{\rm LLL}\cr + \frac12 \phi K_\phi^{-1} \phi + O(N^{-1/2}),\quad K_\phi^{-1}=r_H - \kappa_\phi\nabla^2.
\end{align}
Dropping $O(N^{-\frac12})$ and integrating out $\phi$ generates the Ising channel $-\frac{g^2}{2}\,n^zK_\phi n^z$, however, $K_\phi=(r_H-\kappa_\phi \nabla^2)^{-1}$ is nonlocal. For small $\kappa_\phi$, a reasonable approximation is derivative expansion, followed by static approximation, $K_\phi\approx r_H^{-1} + \kappa_\phi r_H^{-2}\nabla^2 + O(\kappa_\phi^2)\approx r_H^{-1} + \kappa_\phi r_H^{-2} \vec\nabla^2$. This makes $K_\phi$ ultralocal, and gives the desired Hamiltonian:
\begin{equation}
\begin{aligned}
\cH_I^{\rm fuzzy}={}&
-\frac{1}{2}\int\dd\Omega_1\,\dd\Omega_2\, U_z(\Omega_{12})
n^z_{\LLL}(\Omega_1)n^z_{\LLL}(\Omega_2)\cr &+m\int\dd\Omega\,n^x_{\LLL}(\Omega),
\end{aligned}
\label{eq:fuzzyIsing0}
\end{equation}
in which the potential $U_z(\Omega_{12})=\langle\Omega_1|g^2 K_\phi|\Omega_2\rangle$ now matches the one in \cite{ZhuHanHuffmanHofmannHe2023}. Alternatively, we could \emph{define} $K_\phi=g_1 + g_2 \vec\nabla^2$, in which case $K_\phi^{-1}$ in \eqref{eq:intermediate} becomes nonlocal and instantaneous (however, $\partial_\tau^2$ is generated once $\Psi_{\rm LLL}$ is integrated out, see Supplemental Material). 

The symmetries of \eqref{eq:fuzzyIsing0} descending from \eqref{eq:IsingMother} include rotations, Ising  $\Z_2$ flip $n^z_{\rm LLL}\mapsto -n^z_{\rm LLL}$, and $\widetilde{CP}, \widetilde{CT}$ that on the LLL become particle-hole symmetries (PH) \cite{ZhuHanHuffmanHofmannHe2023} (the $S^2$-orientation-reversing unitary one, and orientation-preserving antiunitary one, respectively). Thus the mother theory provides an elegant link:
\begin{equation*}
\boxed{\text{PH symmetries of }\cH_I^{\rm fuzzy}} \leftarrow \boxed{\substack{\widetilde{CP}\\\\ \widetilde{CT}}} \rightarrow \boxed{\text{Parities of } \cL_{\rm loc. EFT}}
\end{equation*} 

\section{Gaussian kernel and correlators}

Correlators of $\phi$ directly flow to the CFT correlators. Integrating out $\phi$ in \eqref{eq:intermediate} is Gaussian and identifies:
\begin{equation}
\phi = - g K_\phi n^z_{\rm LLL}  + \eta,
\end{equation}
where $\eta$ is Gaussian noise of covariance $K_\phi$, so the CFT correlators are not just the $n^z_{\LLL}$ correlators. One must act by $K_\phi$ and include the Gaussian shift, e.g.: 
\begin{equation}
\langle \phi\phi\rangle =g^2 \langle (K_\phi n^z_{\LLL}) (K_\phi n^z_{\LLL}) \rangle + \langle\eta\eta\rangle.
\end{equation}
This is inconvenient for non-local $K_\phi=(r_2 - \kappa_\phi\nabla^2)^{-1}$, but the ultralocal static $K_\phi = g_1 + g_2 \vec\nabla^2$ helps: the Gaussian shift $\langle\eta\eta \rangle=g_1\delta + g_2 \vec\nabla^2\delta$ becomes a contact term that is dropped in the separate-point correlators, and the action by $K_\phi$ relates $n^z_{\LLL}$ to $\phi$ up to descendants, as observed before \cite{HuHeZhu2023}. The ``reverse route'' \emph{defines} $\cL_{int}$ in \eqref{eq:intermediate} with $K_\phi = g_1 + g_2\vec\nabla^2$, $O(N^{-1/2})$ absent, and $\cH_I^{\rm fuzzy}$ exact. The Gaussian kernel $K_\phi^{-1}$ is static and nonlocal in space, yet the $\phi$ dynamics generated by the $\Psi_{\rm LLL}$ loop. This gives a $\phi$ EFT exactly equivalent to $\cH_I^{\rm fuzzy}$ (see Supplemental Materials \cite{SupplementalMaterial}).

\section{Density channel and massive HS field}
The model of \cite{ZhuHanHuffmanHofmannHe2023} also includes a singlet-density channel:
\begin{equation}\label{eq:improvementHn0n0}
\delta\cH_{n^0n^0} = \frac12\int\dd\Omega\,\dd\Omega'\,
U_z(\Omega,\Omega')\delta n^0_{\LLL}(\Omega)\delta n^0_{\LLL}(\Omega'),
\end{equation}
where $\delta n^0_{\LLL}=n^0_{\LLL} - \rho_{\LLL}$, with $\rho_{\LLL}=\frac{N}{4\pi R^2}$. Actually, \cite{ZhuHanHuffmanHofmannHe2023} use $n^0_{\rm LLL}$ not $\delta n^0_{\LLL}$, but for fixed filling, the difference is a harmless constant energy shift. We write $\delta n^0_{\LLL}$ because the Dirac vacuum with mass term $m n^x$ has half-filled LLL, and the properly normal-ordered $\bar\Psi\Psi$ projects to $\delta n^0_{\LLL}$.
The term \eqref{eq:improvementHn0n0} does not change the universality class, it reduces the finite-size drift of \eqref{eq:fuzzyIsing0}. It can also be HS-transformed using a massive non-critical scalar $\rho$:
\begin{equation}
\cL_\rho = \frac12 \rho K_\rho^{-1}\rho + ig\rho \bar\Psi\Psi,
\end{equation}
where $K_\rho=K_\phi$ if \eqref{eq:improvementHn0n0} has the same potential $U_z$ as \eqref{eq:fuzzyIsing0} (like in \cite{ZhuHanHuffmanHofmannHe2023}).
Note that $\bar\Psi\Psi$ is odd under $\widetilde{CP}$ and $\widetilde{CT}$. With $\widetilde{CP}$ unitary and $\widetilde{CT}$ antiunitary, we define:
\begin{equation}
\widetilde{CP}: \rho \mapsto -\rho,\quad \widetilde{CT}: \rho\mapsto \rho,
\end{equation}
ensuring that the full theory is still $\widetilde{CP}$ and $\widetilde{CT}$ invariant.

$\Psi$ has mass eigenvalues $ig\rho \pm \sqrt{m^2 + g^2\phi^2}\neq 0$. Integrating $\Psi$ and $\rho$ out produces the same EFT \eqref{eq:scalarEFT}, but with different coefficients. However, tuning $\phi$ to criticality makes $\rho$ very heavy (of order $\Delta_{\LLL}$), making the recipe ``first integrate $\Psi$, then $\rho$'' unreliable. Instead, we verify its effect numerically, considering
\begin{eqnarray*}
	\cH_I^{\rm fuzzy} + \alpha \delta\cH_{n^0 n^0,}
\end{eqnarray*}
which at $\alpha=1$ reproduces the Hamiltonian of \cite{ZhuHanHuffmanHofmannHe2023}. Conformal perturbation theory (CPT), following \cite{LauchliEtAl2025} (see \cite{SupplementalMaterial} for details), shows: at $\alpha=1$ the leading irrelevant coupling is suppressed by order of magnitute compared to $\alpha=0$ (similar to the choice of $V_0$ having the same effect in \cite{LauchliEtAl2025}). The result of numerical analysis is on the Fig.\ref{fig:alpha-channel}:
\begin{figure}[h]
	\includegraphics[scale=0.5]{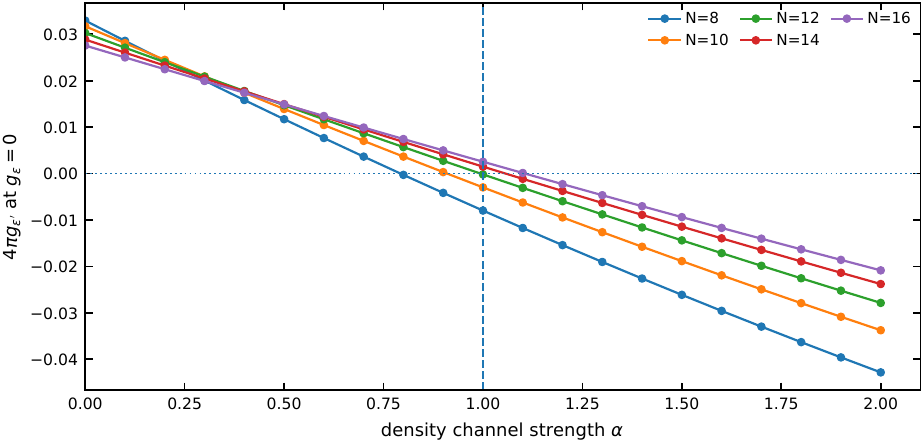}
	\caption{\label{fig:alpha-channel} The leading irrelevant coupling $g_{\varepsilon'}$ at $g_\varepsilon=0$ criticality as a function of $\alpha$. The plots are closer to zero at $\alpha=1$.}
\end{figure}\\
Allowing $K_\rho\neq K_\phi$, tuning their parameters independently and including more states in CPT would enable suppressing more irrelevant couplings.

\section{Same mother theory, different CFT}
A $\Z_2$-invariant EFT of $\phi$ should be tunable to the Gaussian point. In \eqref{eq:scalarEFT} this requires $u_4\to 0$, which, however, is a singular limit as seen from \eqref{eq:scalarEFT-coefficients}. (Actually, we need $u_4$ on fluxed $S^2\times \R$, which is computed in the Supplemental Material \cite{SupplementalMaterial}, but the conclusion is the same.) Thus the simple Ising mother \eqref{eq:IsingMother} has no useful free scalar point. Could adding a massive HS field $\rho$ with $K_\rho$ independent of $K_\phi$ make $u_4\to 0$ possible? Assuming $u_6>0$ and setting $r_2=(4R^2)^{-1}$, this is the tricritical route to free scalar \cite{TaylorEtAl2026}, giving numerical evidence that the answer is yes. Indeed, independently varying $K_\phi$ and $K_\rho$ means separately adjusting interlayer and intralayer potentials, as in \cite{TaylorEtAl2026}. In our language, we integrate out $\Psi$ and $\rho$, and tune $\phi$ to tricriticality by adjusting $m, K_\phi, K_\rho$.

Finally, allowing $\Z_2$-breaking couplings, Lee--Yang (LY) can be reached by an imaginary Ising-odd deformation~\cite{FanDongVishwanath2026,ArguelloCruzEtAl2025,EliasMiroDelouche2025}. In the mother theory, this is simply:
\begin{equation}\label{eq:LY}
\cL_I + \delta\cL_{\rm LY},\quad \delta\cL_{\rm LY} = i \lambda \phi,
\end{equation}
which descends to the same deformation of the EFT and reaches the LY point. Integrating out $\phi$ first produces the fuzzy sphere model from~\cite{FanDongVishwanath2026,ArguelloCruzEtAl2025,EliasMiroDelouche2025}, which is just the Ising Hamiltonian deformed by $i h_z \int \dd\Omega\, n^z_{\rm LLL}(\Omega)$. This exhausts the possibilities for one real $\phi$ and a doublet $\Psi$.

\section{Complex HS field: the easy-plane family}
\label{sec:easy-plane}

The next minimal change is to replace the real mediator by a complex field $\Phi$ carrying flavor charge. Define:
\begin{equation}
 n^+=\bar\Psi\sigma_F^+\Psi=\bar\psi_1\psi_2,
 \qquad
 n^-=\bar\Psi\sigma_F^-\Psi=\bar\psi_2\psi_1,
\label{eq:npm}
\end{equation}
and consider:
\begin{equation}
\begin{aligned}
\cL_{\rm ep}={}&\bar\Psi \gamma^\mu D_\mu[A]\Psi+\mu n^z
+g(\Phi n^-+\Phi^*n^+)\\
&+\Phi^*K_\Phi^{-1}\Phi,
\qquad K_\Phi^{-1}=r_\Phi-\kappa_\Phi\nabla^2.
\end{aligned}
\label{eq:complexMother}
\end{equation}
It has an easy-plane $U(1)_F$ flavor symmetry:
\begin{equation}
\Psi\to e^{i\alpha\sigma_F^z/2}\Psi,
\qquad
\Phi\to e^{-i\alpha}\Phi,
\label{eq:U1ep}
\end{equation}
in addition to the fermion number $U(1)_V$ (coupled to the flux background $A$), under which $\Psi$ has charge $+1$. At zero flux, we also have the charge-conjugation $C$, and parities $\widetilde{P}=R_F\circ P$, $\widetilde{T}=R_F\circ T$. Here $R_F$ is the reflection $\psi_1 \leftrightarrow \psi_2$, $\Phi \leftrightarrow \Phi^*$ (extending $U(1)_F$ to $O(2)_F$); $R_F$, $P$, and $T$ are individually broken by $\mu\neq 0$, while $R_F\circ P$ and $R_F\circ T$ are preserved. Thus the flat space theory preserves $C, \widetilde{P}$, and $\widetilde{T}$.

The gapped fermions in $\R^{3}$ generate a complex scalar EFT constrained by symmetries:
\begin{equation}
	\begin{aligned}
		\cL_\Phi={}&Z_\Phi|\partial\Phi|^2+r_2|\Phi|^2+u_4|\Phi|^4 +\dots\\
	\end{aligned}
	\label{eq:complexEFT}
\end{equation}
The coefficients $r_2, u_4, u_6, \dots$ are the same as in \eqref{eq:scalarEFT-coefficients} (with $m$ replaced by $\mu$), because a $U(1)_F$ rotation makes a constant $\Phi$ real, making the fermion determinant identical to the Ising case (this argument fails for $Z_\Phi$ and other coefficients captured by a non-constant $\Phi$ background). As $u_4>0$, tuning $r_2$ reaches the $O(2)$ Wilson--Fisher CFT. This, however, does not carry over to the fluxed $S^2\times \R$ (which is why the fuzzy-sphere realizations of $O(2)$ require at least a triplet of fermions \cite{GuoEtAl2026,DeyEtAl2026O2}).

On $S^2\times\R$ the magnetic flux only preserves $\widetilde{CP}=C\circ \widetilde{P}$ and $\widetilde{CT}=C\circ \widetilde{T}$, allowing a non-relativistic $C$-odd Berry term in the EFT:
\begin{equation}\label{eq:Berry}
\cL_B=c(\bar\Phi\partial_\tau\Phi - \Phi\partial_\tau\bar\Phi),\quad c=-\operatorname{sgn}(\mu)
\frac{Ng^2}{32\pi R^2\mu^2}.
\end{equation}
The coefficient $c$ is generated by LLL, while paired HLLs cancel (see Supplemental Material \cite{SupplementalMaterial}). 
Berry terms are topological terms in the EFT \cite{Dedushenko:2022pem,Kapustin:2020eby}. Here the underlying topological term is $-\frac{{\rm sgn}(\mu)g^2}{16\pi\mu^2}F\wedge(\bar\Phi\dd\Phi-\Phi\dd\bar\Phi)$, where $F$ is the $U(1)_V$ background field strength. It is perfectly covariant and only becomes Lorentz-violating in the flux background $F=\frac{N}{2 R^2}{\rm Vol}_{S^2}$.

So the leading kinetic term \eqref{eq:Berry} is non-relativistic. The $\cL_\Phi$ EFT has a $U(1)_F$-symmetric (disorder) and a spontaneously broken (SSB) phases. We either: (1) tune to the critical point (it has a non-relativistic $z=2$ scaling); or (2) stay deep inside the SSB phase, where $\Phi=\varrho e^{i\theta}$, a heavy $\varrho$ decouples, the Berry term vanishes for $\theta$ alone, and so the Lorentz symmetry is restored. Both cases are interesting: the former is the first fuzzy realization of $z=2$ criticality (a free Schrodinger-symmetric dilute magnon point); the latter is close to the free relativistic scalar of \cite{He2025FreeScalar} once the conformal mass $\frac{n^x}{R^2}$ is added.

Integrating out HLLs and $\Phi$, followed by our usual approximations, gives the easy-plane Hamiltonian:
\begin{equation}
\begin{aligned}
\cH_{\rm ep}^{\rm fuzzy}={}&
-\frac{1}2\int\dd\Omega\,\dd\Omega'\,
U_\perp(\Omega,\Omega')
\big[n^x(\Omega)n^x(\Omega')\nobreak\\
&+n^y(\Omega)n^y(\Omega')\big]-h_z\int\dd\Omega\,n^z(\Omega).
\end{aligned}
\label{eq:easyPlaneHamiltonian}
\end{equation}
The $U(1)_F$ is generated by $\int\dd\Omega\,n^z(\Omega)$, so $h_z=-\mu$ is a chemical potential. When $h_z=0$, we are in the SSB phase, with parallel spins pointing in some direction in the $xy$ plane. When $h_z$ is large, the ground state is maximally polarized along $z$, indicating disordered phase. As we lower $h_z$, sectors with $k$ magnons (spin flips) are shifted by $(2k-N) \Delta h_z$, leaving their internal dispersion unchanged. The disorder persists until we reach the (critical or saturation) point, at which the magnon gap closes. It is important that all $k$-magnon gaps close simultaneously, not \emph{before} the one-magnon gap (avoiding bound states preemption \cite{SupplementalMaterial}), thus indicating the dilute magnon gas described near the critical point by:
\begin{equation}
	b^* \left(\partial_\tau - \mu_b - \frac{\vec\nabla^2}{2M_*}\right) b + \frac{u}{2} (b^* b)^2 + \dots
\end{equation}
Tuning $\mu_b=2(h_{\rm crit}-h_z)\to 0$ reaches the free Schrodinger point $b^* \left(\partial_\tau - \frac{\vec\nabla^2}{2M_*}\right) b$ that has $z=2$, and $u>0$ is marginally irrelevant in (2+1)D \cite{FisherHohenberg1988,FisherWeichmanGrinsteinFisher1989}. This dilute-boson quantum critical point is of course well-known \cite{SachdevSenthilShankar1994,SachdevDunkel2006,FisherWeichmanGrinsteinFisher1989}, but our fuzzy model \eqref{eq:easyPlaneHamiltonian} is new. Note that the relativistic state-operator correspondence is lost (and replaced by the Schrodinger version \cite{NishidaSon2007,Boisvert:2025hex,Boisvert:2026mgw}), the $S^2$ spectrum is not related to local operators. CFT emerges in the thermodynamic limit $N,R\to\infty, \frac{N}{R^2}={\rm fixed}$. Approximating $U_\perp(\Omega,\Omega')$ by the two pseudopotentials $V_0, V_1$, we can solve the one-magnon sector exactly (see Supplemental Material \cite{SupplementalMaterial}). At $h_{\rm crit}=V_1(4-6/N)$ the one-magnon gap closes, the dispersion law has $z=2$, while the higher-magnon preemption does not happen, corroborating the dilute gas claim. Thus \eqref{eq:easyPlaneHamiltonian} is an interacting fuzzy sphere model with exact solutions and $z=2$ criticality. The numerical evidence is shown in Figure \ref{fig:z2-easy-plane} of the End Matter, and technical derivations -- in \cite{SupplementalMaterial}.

In the ordered phase $\Phi=\varrho e^{i\theta}$, with massive $\varrho$, we get the shift-symmetric Goldsone $\theta$. A conformal model closely related to \cite{He2025FreeScalar} is obtained by picking $h_z=0$ in the broken phase, and adding a pinning term $\sim \frac{\theta^2}{R^2}$ (see the End Matter for a passage from \eqref{eq:easyPlaneHamiltonian} to \eqref{eq:ep-pinned}),
\begin{equation}
	\begin{aligned}
		H_{\rm ep} = \int\dd\Omega \left[(n^0)^2 - U \sum_{a=x,y} n^a\vec\nabla^2 n^a + \frac{h_x}{R^2} n^x\right].
	\end{aligned}
	\label{eq:ep-pinned}
\end{equation}
This differs from He's model \cite{He2025FreeScalar} by $U \int\dd\Omega\, \cO$, where $\cO=\sum_{a=x,y,z}n^a\vec\nabla^2 n^a$. The two models are very close numerically, suggesting that $\cO$ is highly irrelevant. Both can be optimized towards free $\theta$ spectrum, both suffer from contamination by a not fully decoupled $\varrho$. Figure \ref{fig:goldstone-comparison} in the End Matter substantiates the numerical comparison of the two models.

The easy-plane mother contains further interesting possibilities, such as the relevant quadratic Yukawa $\Phi^2 n^+ + (\Phi^*)^2n^-$ breaking $U(1)_F$ to $\Z_3$ that points at the $\Z_3$ Potts. We will investigate it elsewhere.

\section{Outlook}
The new conceptual framework for understanding the origins of the fuzzy sphere method is both transparent and powerful. Starting from a given mother theory, we can study its deformations and how they map to the LLL Hamiltonian. Here we demonstrated it in two examples: the 3D Ising mother (that effortlessly connects to the tricritical realization of the free scalar and the Lee-Yang CFT), and the easy-plane mother theory, which uncovered a surprising link between the other free scalar realization \cite{He2025FreeScalar} and our novel interacting fuzzy model that admits some exact solutions and has a $z=2$ critical point. This brings a satisfying sense of order, and in the companion paper \cite{NonGaugeMotherTheories}, we survey and further explore mother theories for other known fuzzy sphere models from the literature \cite{HanHuZhu2024,ZhouHuZhuHe2024,ChenEtAl2024SO5,YangEtAl2026O4,ZhouHe2025SpN,YangEtAl2025Potts,FanDongVishwanath2026,ArguelloCruzEtAl2025,EliasMiroDelouche2025,He2025FreeScalar,TaylorEtAl2026,GuoEtAl2026,DeyEtAl2026O3,DeyEtAl2026O2,Stergiou2026,ZhouGaiottoHe2025,VoineaEtAl2026Majorana,TangEtAl2026SUSY}, including the $O(N)$, $SO(5)$ deconfined criticality, the $Sp(N)$ family of models, Potts, etc. The models for Chern-Simons-matter theories \cite{ZhouWangHe2025,HuffmanEtAl2026} also admit mother theories, although the presence of gauge symmetry requires a somewhat special treatment that is subject of another manuscript in preparation. We should also note that the mother theory approach is not limited to fuzzy sphere, other spatial geometries are perfectly fine. This may be especially relevant for gapped QFTs. Indeed, the new approach affords better control over the construction without relying on universality. This should allow us build fuzzy models for gapped QFTs (such as 3D Yang-Mills, for example), and for those, fuzzy torus is expected to be more natural.

\section{Acknowledgements}
The author benefited from conversations and/or correspondence with Yin-Chen He and Zheng Zhou (whose FuzzifiED package \cite{Zhou2025FuzzifiED} was also indispensable). These results were first announced at Strings-2026 on July 7, 2026.

\paragraph{Statement on the AI use.} ChatGPT was used to assist with the development of numerical code and with computations. The paper was written by a human, and the ideas are those of the author, who bears full responsibility for everything presented in this work.


\bibliography{fuzzy-literature}

\onecolumngrid
\appendix
\newpage
\section{Exact easy-plane saturation point}
\label{app:saturation}
\begin{figure*}[h]
	\centering
	\includegraphics[width=0.98\textwidth]{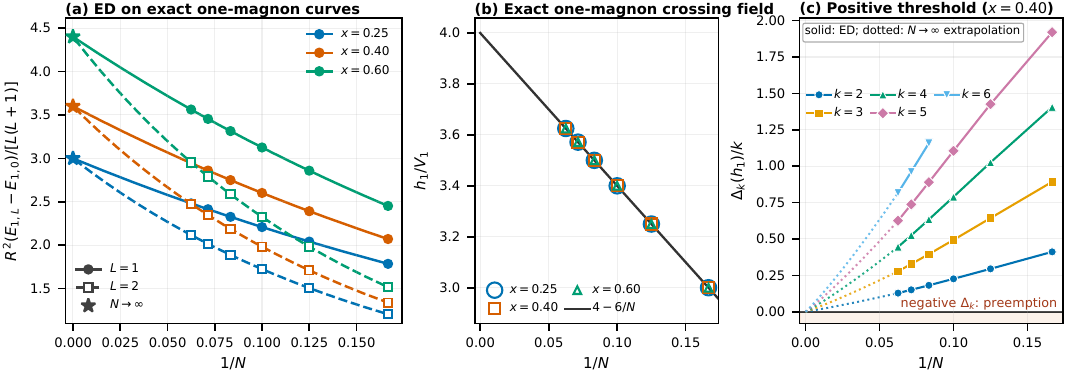}
	\caption{\label{fig:z2-easy-plane}
		The $z=2$ easy-plane saturation transition. We write $x=V_1/(V_0+V_1)$ and fix units such that $V_0+V_1=1$.
		(a) Exact diagonalization (ED) markers for the $L=1,2$ one-magnon energies lie on
		the exact finite-$N$ curves (not fits). Stars at $1/N=0$ mark the
		common limit $R^2(E_{1,L}-E_{1,0})/[L(L+1)]=2+4x$ derived in \cite{SupplementalMaterial}.
		(b) The exact one-magnon crossing field $h_1/V_1=4-6/N$ and three
		coincident data sets displayed with nested open markers. (c) At
		$x=0.40$, the finite-volume multimagnon gap per magnon $\Delta_k(h_1)/k$ is
		positive through $k=6$; a negative value would signal a bound state gap closing first (multiparticle preemption). Dotted tails are extrapolations to the common zero at $1/N=0$ (thermodynamic limit).}
\end{figure*}

\section{Comparison of the two pseudo-Goldstone models}
Firstly, let us show how to go from \eqref{eq:easyPlaneHamiltonian} to \eqref{eq:ep-pinned}. In the two-pseudopotential approximation, $U_\perp(\Omega_1,\Omega_2)= g_1  \delta(\Omega_{12}) + g_2 \vec\nabla^2\delta(\Omega_{12})$. The $\delta(\Omega_{12})$ potential leads to contact interactions that simplify according to:
\begin{equation}
	(n^x)^2 = (n^y)^2 = 2n^0 - (n^0)^2,
\end{equation}
as was noticed in \cite{He2025FreeScalar}. The linear term becomes $\sim \int \dd\Omega\, n^0(\Omega)$, which in a fixed filling sector is just a constant energy shift that can be dropped. The quadratic term becomes $g_1 (n^0)^2$. The $\vec\nabla^2\delta(\Omega_{12})$ potential leads to the $n^x\vec\nabla^2 n^x + n^y \nabla^2 n^y$ interaction. Thus in the end, dropping the constant shift, we get:
\begin{eqnarray}
	\cH^{\rm fuzzy}_{\rm ep} = \int \dd\Omega\, \left[g_1 (n^0)^2 - \frac{g_2}{2}(n^x\vec\nabla^2 n^x + n^y\vec\nabla^2 n^y) - h_z n^z\right].
\end{eqnarray}
Setting $h_z=0$, choosing the energy units such that $g_1=1$, denoting $U=\frac{g_2}{2}$, and adding the pinning term $\frac{h_x}{R^2}n^x$, we obtain \eqref{eq:ep-pinned}. It differs from He's model \cite{He2025FreeScalar} by the irrelevant coupling $\sum_{a=x,y,z}n^a\vec\nabla^2 n^a$, and the numerical results of the two models are very close (with \eqref{eq:ep-pinned}, apparently, giving a slightly better precision), as is shown in Fig. \ref{fig:goldstone-comparison}:
\begin{figure*}[h]
	\centering
	\includegraphics[width=0.98\textwidth]{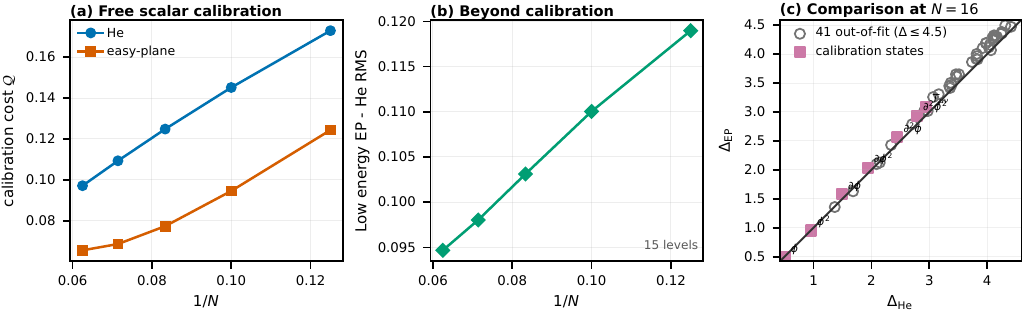}
	\caption{\label{fig:goldstone-comparison}
		Pseudo-Goldstone comparison between easy-plane \eqref{eq:ep-pinned} and He's model \cite{He2025FreeScalar}. (a) Free scalar calibration cost using seven lowest states versus
		$1/N$ for He's published parameters and for the independently optimized
		easy-plane model. (b) RMS easy-plane--He difference for the lowest 15 symmetry-matched
		levels not used in the calibration; the difference steadily decreases with $1/N\to0$. (c) Level by level comparison at the largest size; filled squares denote 7 calibration states and open circles denote 41 lowest noncalibration states. See Supplemental Material \cite{SupplementalMaterial} for the details behind these analyses.}
\end{figure*}

\renewcommand{\thesection}{S\arabic{section}}
\renewcommand{\thesubsection}{\thesection.\arabic{subsection}}
\renewcommand{\theequation}{S\arabic{equation}}
\renewcommand{\thefigure}{S\arabic{figure}}
\renewcommand{\thetable}{S\arabic{table}}

\newpage

\begin{center}
	{\large\bfseries Supplemental Material \label{supplement} for\\[2pt]
		``The Effective Field Theory Origin of the Fuzzy Sphere Method''}
	
	\vspace{0.8em}
	Mykola Dedushenko
	
	\vspace{0.3em}
	{\itshape
		Center for Mathematics and Interdisciplinary Sciences, Fudan University, Shanghai 200433, China\\
		Shanghai Institute for Mathematics and Interdisciplinary Sciences (SIMIS), Shanghai 200433, China}
\end{center}

\vspace{1.2em}

\section{Scope, conventions, and terminology}
\label{sec:scope}

This Supplemental Material contains the details behind technical results invoked in the Letter: the finite-sphere fermion determinant and derivative expansion, the controlled elimination of HLLs, the exact reverse construction starting from the LLL Hamiltonian, the Berry term of the easy-plane mother, exact one-magnon solution and proof that higher-magnon configurations remain gapped when the one-magnon gap closes (and in the thermodynamic limit, all gaps close), and the numerical procedures behind the conformal perturbation theory (CPT), $z=2$ critical point and spectral comparisons.

We work in Euclidean signature on $S^2\times \R$, where the sphere has radius $R$ and supports a $U(1)$ monopole flux $N$. Denote the charged 2D Dirac operator as $\slashed{D}_{S^2}[A]$, so the (2+1)D Dirac operator is $\slashed{D}_{S^2}[A] + \gamma^\tau \partial_\tau$. The spinor LLL, the space of $\slashed{D}_{S^2}[A]$ zero modes, is the spin-$s$ representation,
\begin{equation}
	N=2s+1,
	\qquad
	\cA=4\pi R^2,
	\qquad
	\rho_{\LLL}=\frac{N}{\cA}.
	\label{eq:basic-conventions}
\end{equation}
Notice a $+1$ shift of flux compared to the scalar monopole sphere harmonics used in \cite{ZhuHanHuffmanHofmannHe2023}, for which flux $N-1=2s$ was creating a spin-$s$ representation on the LLL. This shift is of course due to the spin connection on $S^2$.

The nonzero, or higher, Landau levels (HLL) are the eigenstates of $i\slashed{D}_{S^2}[A]$ of paired eigenvalues $\pm\lambda_n$ labeled by $n\geq 1$, which have degeneracies $d_n$ each \cite{Dolan2003DiracSpectrum}, where
\begin{equation}\label{eq:LL-spectrum}
	\lambda_n=\frac{\sqrt{n(N+n)}}{R}, \qquad d_n=N+2n,\quad \text{so} \quad i\slashed{D}_{S^2}[A]|n,\pm,\alpha\rangle = \pm\lambda_n|n,\pm,\alpha\rangle,\ \alpha=1..d_n.
\end{equation}
Energy levels of the (2+1)D Dirac fermion of mass $M$ are eigenvalues of the matrix that, with some particular choice of basis, looks like $\left(\begin{matrix}\lambda_n & M\\ M & -\lambda_n\end{matrix}\right)$; clearly, the pairs of 2D eigenstates $|n,\pm,\alpha\rangle$ get mixed when $M\neq 0$. Thus the HLL energy eigenstates also come in pairs $\Psi_\pm$ related by $\Psi_+ \propto \sigma^y \Psi_-$ (where $\sigma^y$ is the second Pauli matrix), and their energies are given by $\pm E_n(M)$, where:
\begin{equation}
	E_n(M)=\sqrt{\lambda_n^2+M^2}.
\end{equation}
In the usual Dirac vacuum at zero temperature and zero chemical potential, all the negative-energy states are filled. As for the LLL states, they are chiral, i.e. not paired, and have energies:
\begin{equation}
	E_0(M) = M \times {\rm Sign},
\end{equation}
where the sign depends on conventions and orientation (or flux sign), so we fix those by assuming Sign$=+1$. Thus, with such conventions, at $M<0$ the Dirac vacuum has a filled LLL, while at $M>0$ the LLL remains empty.

At fixed cylinder radius the first HLL gap grows as:
\begin{equation}\label{eq:LL-gap}
	\Delta_{\rm LL}R\sim \sqrt N,
\end{equation}
which controls the Wilsonian HLL expansion.

For a prescribed bosonic background $X$, the fermion-induced Euclidean effective action is
\begin{equation}\label{eq:effective-action-def}
	e^{-\Gamma_f[X;A]} = \frac{\int\!\mathcal{D}\Psi\,{\mathcal D}\bar\Psi\, e^{-S_f[\Psi,X;A]}}{\int\!{\mathcal D}\Psi\,{\mathcal D}\bar\Psi\, e^{-S_f[\Psi,0;A]}}.
\end{equation}
For a time-independent and spatially uniform background, at inverse temperature $\beta$,
\begin{equation}\label{eq:Omega-def}
	\Gamma_f[X]=\beta\,\Delta\Omega_f(X) = \beta\cA\,\Delta v_f(X).
\end{equation}
Here $\Delta\Omega_f$ is the total grand potential (equivalently, vacuum energy) shift, while $\Delta v_f=\Delta\Omega_f/\cA$ is the induced effective potential density. EFTs of the Habbard-Stratonovich fields have appeared in the literature, see for example \cite{Moon1995InterlayerCoherence}.

\section{Finite sphere Ising effective field theory}
\label{sec:Ising-finite-sphere}

\subsection{Uniform determinant and scalar self-interactions}

We now evaluate the contributions of LLLs and HLLs to the EFT via fermionic determinants. An earlier reference on effective actions induced by fermionic determinants in magnetic filed is \cite{GusyninMiranskyShovkovy1995}. For the Ising mother, a constant real field $\phi$ gives the flavor mass matrix:
\begin{equation}\label{eq:Mphi}
	m\sigma_F^x+g\phi\sigma_F^z, \quad \text{with eigenvalues } \pm M(\phi), \quad \text{where} \quad M(\phi)=\sqrt{m^2+g^2\phi^2},
\end{equation}
so we have $N$ LLL states of energy $-M(\phi)$ and $N$ LLL states of energy $+M(\phi)$. In the Dirac vacuum, the negative-energy LLL states are occupied, so the LLL is naturally half-filled in the Dirac vacuum at zero temperature and zero chemical potential. The LLL and HLL contributions to the total grand potential (with the $\phi=0$ value subtracted) are:
\begin{align}
	\label{eq:Omega-LLL-Ising}\Delta\Omega_{\LLL}(\phi) &=-N\big[M(\phi)-|m|\big],\\
	\label{eq:Omega-HLL-Ising}\Delta\Omega_{\HLL}(\phi)&=-2\sum_{n=1}^{\infty}d_n\big[E_n(M(\phi))-E_n(|m|)\big]+\Delta\Omega_{\rm ct}.
\end{align}
The factor of two in Eq.~\eqref{eq:Omega-HLL-Ising} accounts for the two flavor mass eigenvalues $\pm M(\phi)$, which have identical nonzero-level energies. The counterterm accounts for the UV divergence of mass, so it subtracts that from the scalar mass but does not affect the finite quartic and higher coefficients.

Define
\begin{equation}
	S_p^{\HLL}(m)=2\sum_{n=1}^{\infty}\frac{d_n}{E_n(|m|)^p},
	\qquad
	S_p^{\rm full}(m)=\frac{N}{|m|^p}+S_p^{\HLL}(m).
	\label{eq:Sp-def}
\end{equation}
Expanding Eqs.~\eqref{eq:Omega-LLL-Ising} and \eqref{eq:Omega-HLL-Ising},
\begin{equation}
	\Delta v_f(\phi)
	=-\frac{g^2}{2\cA}S_1^{\rm full}\phi^2
	+\frac{g^4}{8\cA}S_3^{\rm full}\phi^4
	-\frac{g^6}{16\cA}S_5^{\rm full}\phi^6+\cdots .
	\label{eq:Ising-potential-expansion}
\end{equation}
In the convention
\begin{equation}
	\cL_{\rm eff}\supset
	\frac{r_2}{2}\phi^2+\frac{u_4}{4!}\phi^4+\frac{u_6}{6!}\phi^6+\cdots,
\end{equation}
the finite induced interactions are
\begin{equation}
	\boxed{
		u_{4,f}=\frac{3g^4}{\cA}S_3^{\rm full}>0,
		\qquad
		u_{6,f}=-\frac{45g^6}{\cA}S_5^{\rm full}.}
	\label{eq:u4u6-finite-sphere}
\end{equation}
This is the finite-sphere result used in the paper. In particular, the minimal Ising mother retains a positive quartic coupling after the exact LL sum, so tuning the renormalized quadratic term sends us towards the ordinary Ising Wilson--Fisher point. The negative induced $u_6$ does not destabilize the theory by itself: the exact determinant grows as $M(\phi)^3$ in flat space and the Wilsonian potential also contain higher-order terms. For the tricritical route, an independent massive $\rho$ channel is expected to tune the net quartic coefficient to zero while keeping the full potential stable (and make $u_6>0$).

For $p>2$, $S_p^{\HLL}$ is convergent. At fixed $R$ and large $N$,
\begin{equation}
	S_p^{\HLL}=2R^pN^{1-p/2}\zeta(p/2)+O(N^{-p/2}),
	\label{eq:Sp-asymptotic}
\end{equation}
so HLL-induced local vertices are suppressed by powers of the LL gap. A useful exact representation is obtained by setting
\begin{equation}
	a=1+\frac N2,
	\qquad
	\eta^2=(mR)^2-\frac{N^2}{4},
	\qquad
	\mathcal Z_\nu(a,\eta)=\sum_{k=0}^{\infty}\big[(k+a)^2+\eta^2\big]^{-\nu},
\end{equation}
which gives, for $p>2$,
\begin{equation}
	S_p^{\HLL}=-\frac{4R^p}{p-2}\,\partial_a\mathcal Z_{p/2-1}(a,\eta).
	\label{eq:Sp-zeta}
\end{equation}

\subsection{Frequency and angular momentum dependence}
More generally, the quadratic quantum correction in the thermal effective action due to fermionic integration is
\begin{eqnarray}
	\Gamma^{(2)} = \frac12 \sum_{\omega, L, M} \phi_{\omega, L, M} \Pi(i\omega, L) \phi^*_{\omega,L,M},
\end{eqnarray}
where $\Pi(i\omega, L)$ is the one-loop diagram. Here we assume:
\begin{eqnarray}
	\phi(\tau, \Omega) = \frac1{\sqrt\beta} \sum_{\omega, L, M} \phi_{\omega,L,M} e^{-i\omega \tau} Y_{LM}(\Omega),
\end{eqnarray}
where the $(L,M)$ label spherical harmonics, and $\omega$ are the bosonic Matsubara frequencies:
\begin{equation}
	\omega_n = \frac{2\pi n}{\beta},\quad n\in\mathbb{Z}.
\end{equation}
The fermions are expanded as:
\begin{equation}
	\Psi=\frac1{\sqrt \beta} \sum_\nu \sum_{a\in {\rm LL}} e^{-i\nu \tau} u_a(\Omega) c_{a,\nu},
\end{equation}
where the fermionic Matsubara frequencies are:
\begin{equation}
	\nu_r = \frac{(2r+1)\pi}{\beta}.
\end{equation}
The interaction vertex:
\begin{align}
	S_{\rm int} = g \int\dd\tau\, \dd A\, \phi(\tau, \Omega) \Psi^\dagger \gamma^\tau \sigma_F^z \Psi= \frac{g}{\sqrt{\beta}}\sum_{\omega, L, M} \sum_\nu \phi_{\omega, L, M} c^\dagger_{a,\nu+\omega} c_{b,\nu} V^{LM}_{ab},
\end{align}
where
\begin{equation}
	V^{LM}_{ab} = \int\dd A\, Y_{L,M}(\Omega) u_a^\dagger(\Omega) \gamma^\tau \sigma_F^z u_b(\Omega), \qquad \dd A=R^2\dd\Omega.
\end{equation}
Now the one-loop contribution $\Pi(i\omega, L)$ is equal to:
\begin{equation}
	\Pi(i\omega, L) = \frac{g^2}{2L+1} \sum_{M=-L}^L \sum_{a,b \in {\rm LL}} \frac1{\beta} \sum_\nu G_a(i\nu)G_b(i\nu + i\omega) |V_{ab}^{LM}|^2,
\end{equation}
where the fermionic Matsubara Green function in momentum space is
\begin{eqnarray}
	G_a(i\nu) = \frac1{i\nu - \epsilon_a},
\end{eqnarray}
where $\epsilon_a$ is the energy of fermionic state (Landau level) $a$. The sum over ${\rm LL}$ goes either over the HLL or all Landau levels, depending on the problem we are studying. The sum over Matsubara frequencies is easy to evaluate using:
\begin{equation}
	\frac1{(i\nu - \epsilon_a)(i\nu + i\omega - \epsilon_b)} = \frac1{i\omega + \epsilon_a - \epsilon_b}\left[\frac1{i\nu - \epsilon_a} - \frac1{i\nu + i\omega -\epsilon_b}\right],
\end{equation}
and then recalling that
\begin{equation}
	\frac1{\beta} \sum_{\nu=\frac{\pi}{\beta}(2r+1)} \frac1{i\nu - \epsilon} = n_F(\epsilon) - \frac12,\quad \text{where}\quad n_F(\epsilon) = \frac1{e^{\beta\epsilon} + 1}.
\end{equation}
Thus we obtain:
\begin{equation}
	\Pi(i\omega, L) = \frac{g^2}{2L+1} \sum_{M=-L}^L \sum_{a,b\in{\rm LL}} \frac{n_F(\epsilon_a) - n_F(\epsilon_b)}{i\omega + \epsilon_a - \epsilon_b} |V_{ab}^{LM}|^2.
\end{equation}
We are of course interested in the zero temperature case, where $n_F(\epsilon)=\Theta(-\epsilon)$ is a step function. Then only the terms with $\epsilon_a \epsilon_b <0$ contribute, so the zero-temperature formula is:
\begin{equation}
	\Pi(i\omega, L) = -\frac{2g^2}{2L+1} \sum_{M=-L}^L \sum_{\substack{\epsilon_a<0\\ \epsilon_b>0}} \frac{\Delta_{ba}}{\omega^2 + \Delta_{ba}^2} |V_{ab}^{LM}|^2,
\end{equation}
where $\Delta_{ba}=\epsilon_b - \epsilon_a$, and we sum over the appropriate Landau levels.

For example, the HLL contribution to uniform polarization is:
\begin{equation}\label{eq:HLL-polarization-uniform}
	\Pi_{\HLL}(i\omega,0) =-\frac{8g^2}{\cA}\sum_{n=1}^{\infty}d_n\frac{E_n}{\omega^2+4E_n^2}, \qquad E_n=E_n(|m|).
\end{equation}
Its low-frequency expansion is:
\begin{equation}\label{eq:HLL-frequency-expansion}
	\Pi_{\HLL}(i\omega,0) =-\frac{g^2}{\cA}S_1^{\HLL} +\frac{g^2}{4\cA}S_3^{\HLL}\omega^2 -\frac{g^2}{16\cA}S_5^{\HLL}\omega^4+\cdots,
\end{equation}
and the coefficient of $\omega^2$ is positive, implying a stable induced temporal kinetic term.

For the LLL, define the form factor
\begin{equation}\label{eq:pL-def}
	p_L(N)=\prod_{j=1}^{L}\frac{N-j}{N+j}, \qquad p_0=1.
\end{equation}
The polarization due to LLL fermions running in the loop is:
\begin{equation}\label{eq:LLL-polarization-Ising}
	\Pi_{\LLL}(i\omega,L) =-\frac{4\rho_{\LLL}g^2|m|}{\omega^2+4m^2}\,p_L(N).
\end{equation}
Since
\begin{equation}
	p_L=1-\frac{L(L+1)}{N}+O(N^{-2})
\end{equation}
for fixed $L$, Eq. \eqref{eq:LLL-polarization-Ising} generates
\begin{equation}
	Z_{\tau,\LLL}=\frac{\rho_{\LLL}g^2}{4|m|^3},
	\qquad
	Z_{s,\LLL}=\frac{g^2}{4\pi|m|}+O(N^{-1}).
	\label{eq:LLL-ZsZt}
\end{equation}
The contribution of HLLs is suppressed at the cylinder scale $\omega_{\rm low}\sim R^{-1}$ by
\begin{equation}\label{eq:HLL-control}
	\frac{\omega_{\rm low}}{\Delta_{\rm LL}}=O(N^{-1/2}).
\end{equation}

\section{The exact reverse route from the fuzzy Hamiltonian}
\label{sec:reverse-route}

Rather than deriving the static LLL Hamiltonian as the leading approximation to a relativistic mother, one may \emph{define} the intermediate LLL-plus-mediator theory to be exact (i.e., lacking the $O(N^{-1/2})$ corrections), and recover the scalar EFT by integrating out the LLL fermions.

Denoting the spin-$s$ representation as $W_s$, the LLL-projected one-body operator algebra is
\begin{equation}\label{eq:fuzzy-harmonic-cutoff}
	\operatorname{End}(W_s)=\bigoplus_{L=0}^{2s}W_L=\bigoplus_{L=0}^{N-1}W_L.
\end{equation}
Hence the projected density $n^a$ has modes only for $0\leq L\leq N-1$. Introduce a band-limited real mediator:
\begin{equation}\label{eq:bandlimited-phi}
	\widehat\phi(\tau)=\sum_{L=0}^{N-1}\sum_{M=-L}^{L} \phi_{LM}(\tau)\,\widehat Y_{LM},
\end{equation}
and a static covariance:
\begin{equation}\label{eq:KL-reverse}
	K_L=g_1-g_2\frac{L(L+1)}{R^2}, \qquad K_L>0 \quad (0\leq L\leq N-1),
\end{equation}
corresponding to $K_\phi=g_1+g_2\boldsymbol\nabla^2$. The exact Euclidean action is
\begin{align}\label{eq:reverse-action}
	S_{\rm rev}={}\int\dd\tau\, c^\dagger\left[\partial_\tau+m\sigma_F^x +g\widehat\phi(\tau)\sigma_F^z\right]c
	+\frac12\int\dd\tau\sum_{L=0}^{N-1}\sum_{M=-L}^{L} (-1)^M\phi_{LM}K_L^{-1}\phi_{L,-M}.
\end{align}
Integrating out $\phi$ is an exact Gaussian operation and gives
\begin{equation}\label{eq:reverse-to-H}
	S_{\rm rev}\longrightarrow
	\int\dd\tau\,c^\dagger(\partial_\tau+m\sigma_F^x)c-\frac{g^2}{2}\int\dd\tau\sum_{LM}(-1)^M K_L n^z_{LM}n^z_{L,-M}.
\end{equation}
Thus Eq.~\eqref{eq:reverse-action} is exactly equivalent to the expected fuzzy Hamiltonian. There is no $O(N^{-1/2})$ remainder in this definition.

Integrating out the LLL fermions instead gives the exact band-limited scalar effective action for the quantum field $\phi$:
\begin{align}\label{eq:reverse-effective-action}
	\Gamma_N[\phi]={}\frac12\sum_{\omega LM}(-1)^M \phi_{\omega LM}K_L^{-1}\phi_{-\omega,L,-M}
	-\Tr_{W_s\otimes\mathbb C^2\otimes\tau}  \log\left[\partial_\tau+m\sigma_F^x + g\widehat\phi(\tau)\sigma_F^z\right]
	+\mathrm{const}.
\end{align}
The trace is finite in the spatial indices.

Again, the mediator has a natural spatial ultraviolet cutoff,
\begin{equation}\label{eq:natural-cutoff}
	L_{\max}=N-1,
\end{equation}
because modes above $N-1$ cannot couple to any LLL bilinear. If one starts with an unrestricted field, those higher harmonics remain decoupled Gaussian modes and may be omitted from the minimal exact representation.

Importantly, although the bare kernel in Eq. \eqref{eq:reverse-action} is instantaneous, the LLL determinant generates time derivatives. The exact quadratic part is:
\begin{equation}\label{eq:reverse-quadratic}
	\Gamma_N^{(2)} =\frac12\sum_{\omega LM}(-1)^M\phi_{\omega LM}\left[K_L^{-1}+\Pi_{\LLL}(i\omega,L)\right] \phi_{-\omega,L,-M},
\end{equation}
with $\Pi_{\LLL}$ given in Eq.~\eqref{eq:LLL-polarization-Ising}. At small frequency,
\begin{align}\label{eq:reverse-generated-kinetic}
	K_L^{-1}+\Pi_{\LLL}(i\omega,L)
	={}K_L^{-1}-\frac{\rho_{\LLL}g^2}{|m|}p_L +\frac{\rho_{\LLL}g^2}{4|m|^3}p_L\,\omega^2
	+O(\omega^4).
\end{align}
The positive $\omega^2$ term is the generated kinetic coefficient mentioned in the Letter. Expanding $K_L^{-1}$ and $p_L$ at fixed $L$ also produces a local spatial gradient expansion. Higher even vertices are analytic for external frequencies $|\omega|\ll2|m|$ because the LLL fermions remain gapped. At finite $N$, Eq. \eqref{eq:reverse-effective-action} is a band-limited, generally nonlocal scalar theory; for fixed low $L$ as $N\to\infty$, it approaches the ordinary local Ising EFT.

\section{Density channel and conformal perturbation theory analysis}
\label{sec:CPT}

The density channel scan uses
\begin{equation}\label{eq:CPT-H}
	H_{\alpha,h} =-hN^x+\frac12\sum_{LM}(-1)^M U_L \left[\alpha\,\delta n^0_{LM}\delta n^0_{L,-M}  -n^z_{LM}n^z_{L,-M}\right],
\end{equation}
with fixed $V_0=4.75$, $V_1=1$, sizes $N=8,10,12,14,16$, and $\alpha=0,0.1,\ldots,2$. Thus $\alpha=0$ is the pure Ising channel, while $\alpha=1$ reproduces the relative density and Ising channel normalization of the standard fuzzy Ising Hamiltonian. We keep the interaction kernel $U_L$ fixed throughout the scan so that the observed change can be attributed to the density channel $\alpha$ alone.

\subsection{First order CPT: from gaps to couplings, following \cite{LauchliEtAl2025}}

Near the Ising fixed point, the low-energy Hamiltonian on the cylinder is parameterized, to first order in the leading allowed perturbations, as
\begin{equation}\label{eq:CPT-cylinder-H}
	H_{S^2}=\frac{v}{R}\left[ D+4\pi g_{\epsilon}\widehat\epsilon+4\pi g_{\epsilon'}\widehat\epsilon'+4\pi g_C\widehat C+\cdots\right].
\end{equation}
Here $D$ is the CFT dilatation operator, $v$ is the nonuniversal emergent velocity, $\widehat\epsilon$ and $\widehat\epsilon'$ are the rotational singlets obtained by integrating the corresponding scalar operators over the unit sphere, and $\widehat C$ is the rotationally invariant spherical component of the leading spin-four primary. The couplings in \eqref{eq:CPT-cylinder-H} are dimensionless couplings renormalized at the finite scale $R$. We want to relate them to the microscopic parameters $h$ and $\alpha$.

For a scaling state $|a\rangle$, define $E_a=\langle a|H_{S^2}|a\rangle$. Subtracting the vacuum energy gives:
\begin{equation}\label{eq:CPT-gap-shift}
	e_a\equiv E_a-E_0=\frac{v}{R}\left[\Delta_a+4\pi\sum_{I=\epsilon,\epsilon',C} g_I q_{aI}\right]+O(g_I^2)+\text{higher operators},
\end{equation}
where
\begin{equation}\label{eq:CPT-qdef}
	q_{aI}=\langle a|\widehat{\mathcal O}_I|a\rangle - \langle0|\widehat{\mathcal O}_I|0\rangle.
\end{equation}
For primary states, these diagonal matrix elements reduce to normalized OPE coefficients. For descendants they are fixed by the conformal algebra. Namely, if the external primary has dimension $\Delta_O$ and the perturbing primary has dimension $\Delta_V$ and spin $\ell_V$, define
\begin{equation}\label{eq:CPT-Casimir}
	\mathcal C_2(\Delta_V,\ell_V)=\Delta_V(\Delta_V-3)+\ell_V(\ell_V+1).
\end{equation}
The relative response of the level-one descendant and of the level-two spin-two descendant is, respectively,
\begin{align}
	\label{eq:CPT-A1} A_1(\Delta_O;\Delta_V,\ell_V)&=1+\frac{\mathcal C_2}{6\Delta_O},\\
	\label{eq:CPT-A2} A_{2,2}(\Delta_O;\Delta_V,\ell_V) &=1+\frac{\mathcal C_2^2+10\mathcal C_2(2\Delta_O+1) +2\ell_V(\ell_V+1) [\mathcal C_2-(\ell_V+2)(\ell_V-1)]} {60\Delta_O(\Delta_O+1)}.
\end{align}
These factors determine $q_{\partial\sigma, I}=A_1(\Delta_\sigma; \Delta_I, \ell_I)q_{\sigma, I}$ and $q_{\partial^2_{L=2}\sigma,I}=A_{2,2}(\Delta_\sigma; \Delta_I, \ell_I)q_{\sigma,I}$. 

We use the four gaps ordered as:
\begin{equation}
	(\sigma,\epsilon,\partial\sigma, \partial^2_{L=2}\sigma).
\end{equation}
They obey the linear system
\begin{equation}\label{eq:CPT-linear-system}
	e_a=\mathsf M_{aI}p_I, \qquad p_I=\frac{v}{R} \left(1,\,4\pi g_\epsilon,\, 4\pi g_{\epsilon'},\,4\pi g_C\right),
\end{equation}
where the first column of $\mathsf M$ contains the unperturbed CFT dimensions and the other columns contain the primary OPE coefficients and the descendant factors above, following Ref. \cite{LauchliEtAl2025}. Numerically,
\begin{equation}\label{eq:CPT-response-matrix}
	\mathsf M=
	\begin{pmatrix}
		0.518148806 & 1.051853730 & 0.053040000 & 0.063155200\\
		1.412625280 & 1.532443040 & 1.536200000 & 0.226669714\\
		1.518148806 & 0.293178277 & 0.107235421 & 0.675819718\\
		2.518148806 & 0.146299393 & 0.137072484 & 2.752836180
	\end{pmatrix}.
\end{equation}
Solving Eqn. \eqref{eq:CPT-linear-system} first determines $v/R$ and then the three dimensionless couplings through:
\begin{equation}\label{eq:CPT-couplings}
	4\pi g_\epsilon=\frac{p_2}{p_1},\qquad 4\pi g_{\epsilon'}=\frac{p_3}{p_1},\qquad 4\pi g_C=\frac{p_4}{p_1}.
\end{equation}
This analysis is first order perturbation theory omitting higher order terms and higher irrelevant operators.

\subsection{State identification and critical tuning}

Every $(N,\alpha,h)$ point is independently diagonalized in the symmetry sectors needed below. Whenever the solver returns an arbitrary mixture inside a numerically degenerate energy subspace, we explicitly diagonalize $L^2$ within that subspace to identify the eigenstates. The symmetry assignments are in Table \ref{tab:CPT-states}.
\begin{table}[h]
	\centering
	\begin{tabular}{lccccl}
		\toprule
		state & $\PH$ & $\mathbb{Z}_2$ & $R_y$ & $L$ & sector level \\
		\midrule
		vacuum & $+$ & $+$ & $+$ & 0 & 1 \\
		$\epsilon$ & $+$ & $+$ & $+$ & 0 & 2 \\
		$\sigma$ & $+$ & $-$ & $+$ & 0 & 1 \\
		$\partial\sigma$ & $+$ & $-$ & $-$ & 1 & 1 \\
		$\partial^2_{L=2}\sigma$ & $+$ & $-$ & $+$ & 2 & 1 \\
		\bottomrule
	\end{tabular}
	\caption{Symmetry labels used in the CPT extraction of states. Here $\PH$ is particle--hole parity, $\mathbb{Z}_2$ is the Ising $\mathbb Z_2$, and $R_y$ is spatial rotation by $\pi$ about the $y$ axis evaluated in the $L_z=0$ subspace. ``Sector level'' denotes the energy ordering within a fixed symmetry sector and is not itself a quantum number.}
	\label{tab:CPT-states}
\end{table}

For fixed $(N,\alpha)$, the microscopic transverse field $h$ moves the theory mainly along the relevant direction. We therefore define the finite-size critical field $h_c(N,\alpha)$ by
\begin{equation}\label{eq:CPT-critical-condition}
	g_\epsilon(N,\alpha,h_c)=0.
\end{equation}
Starting from a nearby value of $h$ as a numerical hint, the code searches for a bracket of $h$ values where $g_\epsilon(h)$ changes sign. It then uses a secant step whenever that step lies safely inside the bracket and otherwise bisects. Every trial value of $h$ is re-diagonalized and passed through the state identification and CPT inversion. Once Eqn. \eqref{eq:CPT-critical-condition} is satisfied, the remaining values $g_{\epsilon'}(h_c)$ and $g_C(h_c)$ quantify the leading residual scalar and spin-four deformations. Figure~\ref{fig:alpha-channel} of the Letter displays the signed scalar coupling $g_{\epsilon'}$ along this tuned critical line. The numerical uncertainty for each point is the largest change of the extracted coupling between the selected critical point and the two endpoints of the final $h$ bracket. Its maximum for the plotted $g_{\epsilon'}$ data is about $4.6\times10^{-5}$, far smaller than the markers. It does not include uncertainty in the CFT input data, corrections quadratic in the extracted couplings, or contamination by omitted higher operators. This is why the Letter displays no error bars.

\section{Easy-plane effective action and the induced Berry term}
\label{sec:Berry}

\subsection{Static determinant}

The complex mother is
\begin{equation}\label{eq:complex-mother}
	\cL_{\ep}=\bar\Psi \slashed D[A]\Psi+\mu\bar\Psi\sigma_F^z\Psi +g\left(\Phi\bar\Psi\sigma_F^-\Psi +\Phi^*\bar\Psi\sigma_F^+\Psi\right) +\Phi^*K_\Phi^{-1}\Phi.
\end{equation}
For a uniform $\Phi$, the LLL is a collection of $N$ identical two-level systems,
\begin{equation}\label{eq:two-level-H}
	h_{\LLL}(\Phi)=\mu\sigma^z_F +g\,\mathrm{Re}\Phi\,\sigma^x_F +g\,\mathrm{Im}\Phi\,\sigma^y_F.
\end{equation}
Using the definitions in Eqs.~\eqref{eq:effective-action-def} and \eqref{eq:Omega-def}, the exact LLL grand potential shift and potential density are:
\begin{align}
	\label{eq:complex-Omega} \Delta\Omega_{\LLL}(\Phi) &=-N\left[\sqrt{\mu^2+g^2|\Phi|^2}-|\mu|\right],\\
	\label{eq:complex-v} \Delta v_{\LLL}(\Phi) &=-\rho_{\LLL}\left[\sqrt{\mu^2+g^2|\Phi|^2}-|\mu|\right].
\end{align}
At $\mu=0$, $\Delta v_{\LLL}=-\rho_{\LLL}g|\Phi|$, so the origin is unstable. At nonzero $\mu$,
\begin{equation}\label{eq:complex-static-expansion}
	\Delta v_{\LLL} =-\frac{\rho_{\LLL}g^2}{2|\mu|}|\Phi|^2 +\frac{\rho_{\LLL}g^4}{8|\mu|^3}|\Phi|^4+\cdots.
\end{equation}

\subsection{Exact LLL frequency kernel}

For positive flux in the conventions of the Letter,
\begin{equation}\label{eq:complex-kernel}
	\Pi_{\Phi,\LLL}(i\omega,0)= \rho_{\rm LLL} g^2 \int_{-\infty}^\infty \frac{\dd\nu}{2\pi} \frac{1}{(i\nu-\mu)(i(\nu+\omega)+\mu)} =-\frac{\rho_{\LLL}g^2}{2|\mu|+i\sgn(\mu)\omega}.
\end{equation}
The denominator reflects the spectral asymmetry of the unpaired LLL. Expanding,
\begin{equation}\label{eq:complex-kernel-expansion}
	\Pi_{\Phi,\LLL}(i\omega,0)=-\frac{\rho_{\LLL}g^2}{2|\mu|} +\sgn(\mu)\frac{\rho_{\LLL}g^2}{4\mu^2}i\omega +\frac{\rho_{\LLL}g^2}{8|\mu|^3}\omega^2+\cdots.
\end{equation}
The odd term is first order in time. In the Euclidean convention of the Letter,
\begin{equation}
	\cL_B=c\left(\Phi^*\partial_\tau\Phi-\Phi\partial_\tau\Phi^*\right).
	\label{eq:Berry-convention}
\end{equation}
Matching Eq.~\eqref{eq:complex-kernel-expansion} therefore gives:
\begin{equation}\label{eq:Berry-c}
	\boxed{ c=-\sgn(\mu)\frac{\rho_{\LLL}g^2}{8\mu^2}=-\sgn(\mu)\frac{Ng^2}{32\pi R^2\mu^2}.}
\end{equation}

The same result has a geometric interpretation. Writing
\begin{equation}
	\bm d=(g\,\mathrm{Re}\Phi,g\,\mathrm{Im}\Phi,\mu),
	\qquad h_{\LLL}=\bm d\cdot\bm\sigma,
\end{equation}
the occupied state has the usual two-level Berry curvature \cite{Berry1984} on $\bm d$ space. Pulling it back to the $\Phi$ plane and expanding at small $|\Phi|$ gives a Berry curvature proportional to $\sgn(\mu)g^2/(4\mu^2)$, equivalent to \eqref{eq:Berry-c}. Our computations are also analogous to fermionic determinants appearing in \cite{AbanovWiegmann2000}.

\subsection{The HLL contribution is frequency even}

Every nonzero Landau level has a charge-conjugate partner. The two contribute denominators:
\begin{equation}\label{eq:HLL-even}
	\frac{1}{2E_n-i\omega}+\frac{1}{2E_n+i\omega}=\frac{4E_n}{4E_n^2+\omega^2},
\end{equation}
which are exactly even in $\omega$. Therefore HLLs renormalize $|\Phi|^2$, $|\partial_\tau\Phi|^2$, and higher even-derivative terms, but do not change $c$. The Berry coefficient is fixed entirely by the unpaired LLL.

\subsection{Covariant form of the Berry term}

In the normalization of the Letter, $F=(N/2 R^2)\operatorname{Vol}_{S^2}$. The local term in \eqref{eq:Berry-convention} is the specialization of:
\begin{equation}\label{eq:covariant-Berry}
	\boxed{ \Gamma_{\rm top}=-\sgn(\mu)\frac{g^2}{8\mu^2} \int_{S^2\times\R} \frac{F}{2\pi}\wedge\left(\Phi^*\dd\Phi-\Phi\dd\Phi^*\right).}
\end{equation}
Equation \eqref{eq:covariant-Berry} is diffeomorphism invariant. A fixed magnetic background $F$ converts it into a term with a single time derivative and thereby prevents restoration of Lorentz invariance in the EFT.

There is no analogous physical Berry term for the real Ising mediator. Its mass vector $(m,0,g\phi)$ leads to no Berry connection for the two-level system. Moreover, $F\wedge\dd\phi=\dd(\phi F)$ for a smooth electromagnetic background satisfying $\dd F=0$, so any such single real scalar term is a boundary term even for spatially nonuniform flux.

\subsection{Broken phase}

Writing $\Phi=(v+\sigma)e^{i\theta}$,
\begin{equation}\label{eq:Berry-broken}
	\cL_B=2ic(v+\sigma)^2\partial_\tau\theta.
\end{equation}
The $v^2\partial_\tau\theta$ term is a total derivative in the smooth sector. The remaining $\sigma\partial_\tau\theta$ mixing generates an ordinary $(\partial_\tau\theta)^2$ term after the massive radial mode is integrated out. Thus, while the saturation transition is nonrelativistic due to $\cL_B$, the Goldstone mode $\theta$ deep in the ordered phase can have a relativistic Gaussian action.

\section{Exact easy-plane solutions}
\label{sec:magnons}

Here we describe the one-magnon and two-magnon exact solutions of the easy-plane model, which are enough to locate and test the proposed $z=2$ critical point. We make no claims about the solvability of the full spectrum. The one-magnon problem has one copy of every orbital angular momentum $L=0,\ldots,N-1$ and is diagonalized completely by symmetry, giving a closed finite-$N$ formula. The two-magnon problem has multiplicities, but representation theory reduces it to a finite Racah matrix built from $6j$ and $9j$ symbols.  No diagonalization in the large many-body Hilbert space is required.

\subsection{Hamiltonian, conventions, and the role of magnetic field}
The LLL contains $N=2s+1$ orbitals transforming in the spin-$s$ representation $W_s$, further tensored by the appropriate flavor space. A general two-particle interaction acts on two-particle states, whose orbital part is described, roughly, by $W_s\otimes W_s$. Here we say ``roughly'' because the precise statement depends on the flavor part. In the Ising case, the interaction $\sim U(\Omega_{12})n_\downarrow(\Omega_1) n_\uparrow(\Omega_2)$ is between two different flavors (sometimes called layers), so the orbital part is indeed $W_s\otimes W_s$. More generally, if we have identical particles interacting (such as $n_\uparrow(\Omega_1)n_\uparrow(\Omega_2)$), we might have to take symmetric or antisymmetric square of $W_s$. Now in the two-flavor case, since $W_s\otimes W_s \cong \oplus_{J=0}^{2s} W_J$, the general rotationally invariant two-particle potential is decomposed into irreducible sectors labeled by $J$, and within each sector, due to Schur lemma, the potential is simply a constant, one of the Haldane pseudopotentials. Thus the total interaction is $\sim \sum_{J=0}^{2s} V_r P_r$, where $P_r$ acting on any two-particle state projects it on the subspace of relative angular momentum $r$ (or equivalently total pair angular momentum $J_r=2s-r$); $V_r$ are Haldane pseudopotentials \cite{Haldane1983}. So the most general rotationally invariant two-particle interaction on the LLL, either local or not, is written in terms of $V_0, V_1, \dots, V_{2s}$. We can write $P_r$ explicitly in terms of the orbital modes and real Clebsch-Gordan coefficients as:
\begin{align}
	P_r &= \sum_{m_1,m_2,m_3,m_4}\sum_{M=-J_r}^{J_r} C^{J_r M}_{sm_1,sm_2} C^{J_r M}_{sm_3,sm_4} c^\dagger_{m_1} c^\dagger_{m_2} c_{m_4} c_{m_3}\cr
	&= \sum_{\substack{m_1,m_2,m_3,m_4\\m_1+m_2=m_3+m_4}} C^{J_r (m_1+m_2)}_{sm_1,sm_2} C^{J_r (m_3+m_4)}_{sm_3,sm_4} c^\dagger_{m_1} c^\dagger_{m_2} c_{m_4} c_{m_3}
\end{align}
connecting to familiar equations, say from \cite{ZhuHanHuffmanHofmannHe2023}. In such notations, the normal-ordered easy-plane Hamiltonian is:
\begin{equation}\label{eq:EP-first-quantized}
	H_{\ep} = \sum_{r\geq0} H_r -h_z\sum_m c^\dagger_{m}\sigma_F^z c_m,\quad \text{where}\quad H_r=2(-1)^r V_r \sum_{m_{1,2,3,4},M}C_{m_1m_2}^{J_r M} C_{m_3m_4}^{J_r M} c^\dagger_{m_1\downarrow}c^\dagger_{m_2\uparrow}c_{m_4\uparrow}c_{m_3\downarrow}.
\end{equation}
In a fixed-filling sector, the normal ordered Hamiltonian differs from the non-normal ordered one by a constant shift that does not affect physics and may be ignored.

The $V_r>0$ model is ferromagnetic in the easy plane. The flavor Pauli matrices take usual form, so a flip $\uparrow\mapsto\downarrow$ changes $c^\dagger\sigma_F^zc$ by $-2$; consequently the field $h_z$ raises the energy of every $k$-magnon state by $2kh_z$ relative to the fully $+z$-polarized state. Let such fully polarized state be:
\begin{equation}\label{eq:F-state}
	|F\rangle=\prod_{m=-s}^{s}c_{m\uparrow}^{\dagger}|0\rangle.
\end{equation}
It obeys:
\begin{equation}
	H_r |F\rangle =0.
\end{equation}
so its energy is 
\begin{equation}
	E_F(h_z) = -N h_z.
\end{equation}
If $E_{k,\alpha}{(h_z)}$ denotes some energy level with $k$ flips, then the gap to polarized state is:
\begin{equation}\label{eq:field-shift}
	E_{k,\alpha}(h_z)-E_F(h_z)=E_{k,\alpha}{(0)}+2kh_z.
\end{equation}
Clearly $h_z$ does not change the eigenvectors or the dispersion inside a fixed-$k$ sector. It only moves sectors of different magnon number relative to one another. If $E_{k,0}(0)$ is the zero-field energy of the lowest state in the $k$-magnon sector, it crosses the polarized state $|F\rangle$ (i.e. the gap \eqref{eq:field-shift} closes) at the crossing field $h_k$:
\begin{equation}
	h_k=-\frac{E_{k,0}{(0)}}{2k}. \qquad \boxed{\text{Define the critical field } h_{\rm crit}=\max_{k\geq1}h_k.}
	\label{eq:hk-hsat}
\end{equation}
The dilute magnon transition occurs when the maximum is attained at $k=1$, i.e., $h_{\rm crit}=h_1$.

\subsection{One magnon solution from symmetry}

A one-magnon state contains one particle of flavor polarization $|\downarrow\rangle$ and one hole in $|\uparrow\rangle$. 
Similarly to the computations in \cite{NakajimaAoki1994}, a normalized one-magnon state of total orbital angular momentum $(L,M)$ is:
\begin{equation}\label{eq:one-magnon-state}
	|LM\rangle =\sum_{m_e,m_h}(-1)^{s-m_h} C^{LM}_{s m_e,s,-m_h} c_{m_e\downarrow}^{\dagger}c_{m_h\uparrow}|F\rangle, \qquad L=0,\ldots,2s.
\end{equation}
The orbital Hilbert space is:
\begin{equation}\label{eq:one-magnon-decomposition}
	W_s\otimes W_s^* \simeq W_s\otimes W_s =\bigoplus_{L=0}^{2s}W_L.
\end{equation}
Every $W_L$ occurs exactly once. Hence Schur's lemma proves that every $|LM\rangle$ is an eigenstate of any rotationally invariant one-magnon Hamiltonian, and that its energy depends only on $L$. Finding this energy is a matter of a simple computation. Denote:
\begin{eqnarray}
	|a,b\rangle = c^\dagger_{a\downarrow}c_{b\uparrow}|F\rangle.
\end{eqnarray}
Then:
\begin{eqnarray}
	H_r |a,b\rangle = 2(-1)^r V_r \left[\sum_{m_1,m_2,M} C^{J_r M}_{s m_1, s m_2} C^{J_r M}_{s a, s m_2}|m_1,b\rangle - \sum_{s m_1, s m_2,M}C_{s m_1, s b}^{J_r M} C_{s a, s m_2}^{J_r M}|m_1,m_2\rangle\right].
\end{eqnarray}
Using identities, including the Racah relation to $6j$ symbols:
\begin{align}
	\sum_{n,M} C_{s m, s n}^{J_r M} C_{s p, s n}^{J_r M} &= \frac{2J_r +1}{N}\delta_{mp},\\
	\sum_{m_{1,2}, Q} C_{s n, s m_2}^{J_r Q}C_{s m_1, s q}^{J_r Q} C_{s m_1, s -m_2}^{L M}(-1)^{s-m_2} &= (2J_r +1) \left\{ \begin{matrix}s & s & J_r\\ s & s & L\end{matrix} \right\} C_{s n,s -q}^{LM}(-1)^{s-q},
\end{align}
we then obtain contribution of a single pseudopotential to the energy:
\begin{align}
	&H_r |LM\rangle = E_L^{(r)}|LM\rangle,\\
	\label{eq:one-magnon-6j}
	&\boxed{E_L^{(r)}=2(-1)^r (2J_r+1)V_r \left[\frac1N - \begin{Bmatrix}s&s&J_r\\ s&s&L \end{Bmatrix}\right].}
\end{align}

This equation is the exact solution for an arbitrary rotationally invariant pseudopotential expansion. We only have $V_0$ and $V_1$, and for $r=0,1$, the formula simplifies because $J_r$ is maximal or next-to-maximal.  Define again:
\begin{equation}\label{eq:pL-magnon}
	p_L(N)=\prod_{j=1}^{L}\frac{N-j}{N+j}=\frac{(N-1)!\,N!}{(N-L-1)!(N+L)!}, \qquad p_0=1.
\end{equation}
Then
\begin{align}
	\label{eq:6j-r0}\begin{Bmatrix}s&s&2s\\s&s&L\end{Bmatrix}&=\frac{p_L}{N},\\
	\label{eq:6j-r1}\begin{Bmatrix}s&s&2s-1\\s&s&L\end{Bmatrix}&=\left[\frac{2L(L+1)}{N-1}-1\right]\frac{p_L}{N}.
\end{align}
Substitute this into \eqref{eq:one-magnon-6j} and sum over $r=0,1$ to get the complete one-magnon energy at $h_z=0$:
\begin{equation}\label{eq:one-magnon-exact}
	\boxed{ E_{1, L}(0)=\frac{2(2N-1)V_0}{N}(1-p_L)-\frac{2(2N-3)V_1}{N} \left[1-\left(\frac{2L(L+1)}{N-1}-1\right)p_L\right].}
\end{equation}
For the uniform, or $L=0$, mode the $V_0$ term vanishes because $p_0=1$, so we get:
\begin{equation}\label{eq:e0}
	E_{1,0}(0)=-\frac{4(2N-3)}{N}V_1.
\end{equation}
Hence the exact field at which the uniform one-magnon state crosses the polarized vacuum is:
\begin{equation}\label{eq:h1}
	\boxed{h_1=-\frac{E_{1,0}(0)}{2}=V_1\left(4-\frac6N\right).}
\end{equation}
This is a one-particle statement.  Equality $h_{\rm crit}=h_1$ additionally requires that no $k\geq2$ cluster cross at a larger field, which will be proven in the next subsection.

At $h_z=h_1$, the $L=0$ magnon has zero energy and the exact dispersion relative to it is:
\begin{equation}\label{eq:one-magnon-disp-exact}
	\begin{split}
		\varepsilon_L&=E_{1,L}(0)-E_{1,0}(0)\\
		&=\frac{2(2N-1)V_0}{N}(1-p_L) +\frac{2(2N-3)V_1}{N}\left[1+\left(\frac{2L(L+1)}{N-1}-1\right)p_L\right].
	\end{split}
\end{equation}
In particular,
\begin{align}
	\label{eq:eps1} \varepsilon_1&=\frac{4(2N-1)V_0+12(2N-3)V_1}{N(N+1)},\\
	\label{eq:eps2}\varepsilon_2&=\frac{12(2N-1)V_0}{(N+1)(N+2)}+\frac{12(3N-4)(2N-3)V_1}{N(N+1)(N+2)}.
\end{align}
These are the exact curves plotted in the Letter's Figure~\ref{fig:z2-easy-plane}(a). For fixed $L$,
\begin{equation}\label{eq:pL-largeN}
	p_L=1-\frac{L(L+1)}{N}+\frac{L^2(L+1)^2}{2N^2}+O(N^{-3}).
\end{equation}
We can then write:
\begin{equation}\label{eq:z2-dispersion}
	\boxed{\varepsilon_L = \frac{k_L^2}{2M_*} +O(R^{-4}), \qquad  k_L^2 = \frac{L(L+1)}{R^2}, \qquad M_*=\frac{N}{8R^2(V_0+3V_1)}.}
\end{equation}
The correct thermodynamic limit (analogous to taking a lattice to infinite volume, while holding the microscopic couplings and lattice spacing fixed, and studying the long-wavelength modes) is to send:
\begin{equation}\label{eq:thermo-limit}
	N\to\infty,\qquad R\to \infty,\qquad \frac{N}{2R^2} = \text{fixed magnetic field } B.
\end{equation}
Thus at the long wavelengths, $\varepsilon=p^2/(2M_*)$ with $M_*=\frac{B}{4(V_0+3V_1)}$, proving the $z=2$ dispersion relation for magnons. Of course $\varepsilon_L$ is independent of $h_z$; the field only translates the whole one-magnon branch vertically by $2(h_z-h_1)$. Historical references with similar computations include \cite{BatyevBraginskii1984,SachdevSenthilShankar1994}.

For comparison with Fig.~\ref{fig:z2-easy-plane}, let $x=\frac{V_1}{V_0+V_1}$, and
$\mathcal D_L=\frac{R^2\varepsilon_L}{(V_0+V_1) L(L+1)}$.  Equations \eqref{eq:eps1} and \eqref{eq:eps2} become:
\begin{align}
	\label{eq:D1}\mathcal D_1(N,x)&=\frac{(N-1)[(2N-1)+4x(N-2)]}{N(N+1)},\\
	\label{eq:D2} \mathcal D_2(N,x)&=(N-1)\left[ \frac{(2N-1)(1-x)}{(N+1)(N+2)} +\frac{(3N-4)(2N-3)x}{N(N+1)(N+2)}\right],
\end{align}
with the common limit $\mathcal D_L\to2+4x$ as $N\to\infty$. Exact diagonalization agrees with these to better than $7\times10^{-15}$.

\subsection{Higher magnons and absence of preemption}
\label{sec:ep_allk}

We now prove $h_{\rm crit}=h_1$ for the $V_0,V_1\ge0$ model at fixed
filling $N_\uparrow+N_\downarrow=N$, with $N\ge2$. The proof applies to
every magnon number and any total orbital angular momentum.

Using the same real Clebsch-Gordan convention as above, introduce:
\begin{equation}\label{eq:ep_allk_pairs}
	A_{rq}=\sum_{a,b}C^{J_rq}_{sa,sb}\,c_{b\uparrow}c_{a\downarrow},
	\qquad
	\mathcal C_r=\sum_q A_{rq}^\dagger A_{rq}.
\end{equation}
Equation \eqref{eq:EP-first-quantized} gives
$H_r=2(-1)^rV_r\mathcal C_r$. In particular, the contact contribution
is positive in every magnon sector:
\begin{equation}\label{eq:ep_allk_contact}
	\langle\psi_k|H_0|\psi_k\rangle=2V_0\sum_q\|A_{0q}|\psi_k\rangle\|^2\ge0.
\end{equation}
Now also introduce the adjoints of the one-magnon creation operators in \eqref{eq:one-magnon-state}:
\begin{equation}\label{eq:ep_allk_multipoles}
	B_{LM}=\sum_{a,b}(-1)^{s-b}C^{LM}_{sa,s\,(-b)} c^\dagger_{b\uparrow}c_{a\downarrow}, \qquad B^\dagger_{LM}|F\rangle=|LM\rangle.
\end{equation}
Anticommuting the operators in $\mathcal{C}_r$, such as
$c^\dagger_{b\uparrow}c_{d\uparrow}
=\delta_{bd}-c_{d\uparrow}c^\dagger_{b\uparrow}$, leads to the calculation similar to the one-magnon case, where we use the Racah relation between the Clebsch-Gordan coefficients and the $6j$ symbols again:
\begin{equation}\label{eq:ep_allk_recoupling}
	\mathcal C_r=\frac{2J_r+1}{N}N_\downarrow -(2J_r+1)\sum_{LM} \begin{Bmatrix}s&s&J_r\\s&s&L\end{Bmatrix}	B^\dagger_{LM}B_{LM}.
\end{equation}
This operator identity is general, it holds in every sector.

Applying \eqref{eq:ep_allk_recoupling} to $\mathcal{C}_0$ and $\mathcal{C}_1$, and using expressions \eqref{eq:6j-r0} and \eqref{eq:6j-r1} gives:
\begin{equation}\label{eq:ep_allk_elimination}
	\mathcal C_1=\frac{2(2N-3)}{N}N_\downarrow-\frac{2N-3}{2N-1}\mathcal C_0	-\frac{2(2N-3)}{N(N-1)} \sum_{LM}p_LL(L+1)B^\dagger_{LM}B_{LM}.
\end{equation}
Consequently, with $\widetilde V_0=V_0+\frac{2N-3}{2N-1}V_1$ and $h_1$
as in \eqref{eq:h1},
\begin{align}\label{eq:ep_allk_positive}
	H_{\ep}(h_1)-E_F(h_1)&=H_0+H_1+2h_1N_\downarrow = 2V_0 \mathcal{C}_0 - 2V_1\mathcal{C}_1 + 2V_1\left(4-\frac{6}{N}\right)N_\downarrow\cr
	&=2\widetilde{V}_0\mathcal{C}_0 +\frac{2h_1}{N-1}\sum_{L=1}^{N-1}\sum_M p_LL(L+1)B^\dagger_{LM}B_{LM}\ge 0.
\end{align}
Every coefficient is nonnegative, so equation \eqref{eq:ep_allk_positive} implies:
\begin{equation}\label{eq:ep_allk_gap_bound}
	E_{k,\alpha}(0)\ge-2kh_1, \qquad E_{k,\alpha}(h_z)-E_F(h_z)\ge2k(h_z-h_1).
\end{equation}
Thus every crossing field in \eqref{eq:hk-hsat} obeys $h_k\le h_1$, which becomes equality for the uniform one-magnon state, proving:
\begin{equation}
	\boxed{h_{\rm crit}=h_1=V_1\left(4-\frac6N\right).}
	\label{eq:ep_allk_critical}
\end{equation}
As we lower $h_z$ from the symmetric phase $h_z\gg 0$ towards the ordered or SSB phase, no two-magnon or higher-magnon level can cross $E_F$ before the uniform one-magnon state, which we refer to as the absence of preemption. This result assumes precisely the $r=0,1$ interactions; additional pseudopotentials require a separate analysis.

\subsection{Finite-size gaps and the large-$N$ limit}
\label{sec:ep_finite_size}

Now show that for $V_1>0$, all $k\ge2$ gaps at $h_1$ are strictly positive at
finite $N$. Clebsch-Gordan completeness gives:
\begin{equation}\label{eq:ep_allk_completeness}
	\sum_{LM}B^\dagger_{LM}B_{LM}=N_\downarrow(N-N_\uparrow)=k^2,\qquad B_{00}=\frac{S^+}{\sqrt N},
\end{equation}
where $S^+=\sum_m c^\dagger_{m\uparrow}c_{m\downarrow}$ and $S^-=(S^+)^\dagger$.
The total flavor spin satisfies $S\le N/2$ and $S^z=N/2-k$, so $S^-S^+\le k(N-k+1)$. Hence, in the $k$-magnon sector,
\begin{equation}\label{eq:ep_allk_strict}
	\sum_{L\ge1,M}B^\dagger_{LM}B_{LM}=k^2-\frac{S^-S^+}{N} \ge\frac{k(k-1)(N+1)}{N}>0 \qquad(k\ge2).
\end{equation}
All nonzero-$L$ weights in \eqref{eq:ep_allk_positive} are strictly positive, which together with \eqref{eq:ep_allk_strict} proves the claim.

All these strictly positive gaps, as one may expect, close in the large-$N$ limit. We will show it by establishing an upper bound using the variational principle. Consider the following normalized state:
\begin{equation}\label{eq:ep_allk_dicke}
	|D_k\rangle=\sqrt{\frac{(N-k)!}{k!\,N!}}\,(S^-)^k|F\rangle.
\end{equation}
It obeys $\mathcal C_0|D_k\rangle=0$. We also find that:
\begin{align}
	\langle D_k|\mathcal C_1|D_k\rangle	&=\frac{2(2N-3)k(N-k)}{N(N-1)},\\
	\label{eq:ep_allk_dicke_energy}\langle D_k|H_{\ep}(h_1)-E_F(h_1)|D_k\rangle &=\frac{2h_1k(k-1)}{N-1}.
\end{align}
Denote the lowest $k$-magnon gap at $h_1$ by:
\begin{equation}\label{eq:ep_allk_delta}
	\delta_k(N)=E_{k,0}(0)+2kh_1,
\end{equation}
which, due to the variational principle, is bounded by \eqref{eq:ep_allk_dicke_energy} from above. Together with the lower bound, we get: 
\begin{equation}\label{eq:ep_allk_two_sided}
	\boxed{0<\delta_k(N)\le\frac{2h_1k(k-1)}{N-1}},	\qquad 2\le k\le N,\quad V_1>0.
\end{equation}
Also note that the higher-sector crossing fields satisfy $h_k=h_1-\frac{\delta_k(N)}{2k}<h_1$.

As we send $N\to\infty$ and hold $V_1$ and $k$ fixed, the inequality \eqref{eq:ep_allk_two_sided} implies that $\delta_k(N)\to 0$. Thus indeed, all gaps close in the thermodynamic limit at the critical field $h_z=h_1$, while for finite $N$, only the uniform one-magnon mode becomes gapless at $h_1$. This behavior is confirmed numerically in the Fig.~\ref{fig:z2-easy-plane}(c) of the Letter.

Thus we obtain a gas of free magnons with the dynamical exponent $z=2$ at the critical point $h_1$ in the thermodynamic limit. Recall that the phase transition at $h_1$ only emerges in the thermodynamic limit, and so does the SSB at $h_z\approx0$. Indeed, at finite $N$ and volume, for $h_z=0$, the exact vacuum is symmetric and sits in the $S^z=0$ sector. Also notice that for $h_z\ll 0$, our system lands in the symmetric phase again, with the vacuum $|-F\rangle$ polarized in the opposite $-z$ direction. The flavor reflection symmetry $R_F$ suggests that the respective phase transition must take place at $h_z=-h_1$, and the phase diagram is symmetric under $h_z\mapsto -h_z$. The SSB phase lives in the region $-h_1 < h_z < h_1$, though strictly speaking we have not proven that no other phase transition happens inside. The phase diagram:
\begin{center}
	\begin{tikzpicture}[x=0.9cm, font=\small]
		\draw[->] (-4,0) -- (4,0) node[right] {$h_z$};
		
		\fill (-1.4,0) circle (1.6pt) node[below=5pt] {$-h_1$};
		\fill ( 1.4,0) circle (1.6pt) node[below=5pt] {$h_1$};
		
		\node[above=7pt, align=center] at (-2.7,0)
		{Symmetric\\$-z$ polarized};
		\node[above=7pt, align=center] at (0,0)
		{SSB\\$U(1)_F$ broken};
		\node[above=7pt, align=center] at (2.7,0)
		{Symmetric\\$+z$ polarized};
	\end{tikzpicture}
\end{center}

\subsection{Numerical implementation and checks}
\label{sec:ep_numerics}

Numerical exact diagonalization (ED) of $H_\ep$ is used to obtain the finite-size spectra for $N=6,8,10,12,14,16$ seen in Fig.~\ref{fig:z2-easy-plane} of the Letter. In particular, Fig.~\ref{fig:z2-easy-plane}(a) shows agreement between the numerical and analytical (\eqref{eq:D1} and \eqref{eq:D2}) one-magnon dispersion. The higher-magnon gaps for $k=2,3,4,5,6$ are shown in Fig.~\ref{fig:z2-easy-plane}(c) (the $k=6$ results are only shown for $N=12,14,16$). We use the Julia package \texttt{FuzzifiED} \cite{Zhou2025FuzzifiED} to construct and diagonalize $H_\ep$. Its normal-ordered interaction builder
\texttt{GetDenIntTerms} constructs the two pseudopotential components, with flavor matrices $\sigma_F^x,\sigma_F^y$ and overall coefficient $-1/2$; these components are combined with
weights $V_0,V_1$, implementing \eqref{eq:EP-first-quantized}. Fermionic signs are handled by the package, which places the two flavors consecutively within each orbital. The occupation configurations and basis are constructed using \texttt{Confs} and \texttt{Basis}, fixing
\begin{equation}\label{eq:ep_num_sectors}
	N_{\mathrm e}=N,\qquad L_z=0,\qquad	q_z=N_\uparrow-N_\downarrow=N-2k.
\end{equation}
Every orbital multiplet has integer $L$, so the $L_z=0$ block contains a representative of every multiplet and suffices to determine the minimum energy at fixed $k$. At each coupling, the $h_z=0$ Hamiltonian matrix is constructed with \texttt{Operator} and \texttt{OpMat}. Small blocks are diagonalized as dense Hermitian matrices; larger blocks use the sparse eigensolver \texttt{GetEigensystem}. The latter is initialized with a random vector for a new sector and subsequently with its previously computed lowest state. Total angular momentum is obtained from \texttt{GetL2Terms}. Within numerically degenerate energy subspaces, $\mathbf L^2$ is diagonalized before assigning $L$ from its eigenvalue $L(L+1)$. The calculation obtains the polarized energy $E_F(0)$ and the lowest zero-field energy in each sampled $k$-sector. It then computes:
\begin{equation}\label{eq:ep_num_observables}
	h_k=\frac{E_F(0)-E_{k,0}(0)}{2k},\qquad	\varepsilon_L=E_{1,L}(0)-E_{1,0}(0),\qquad \delta_k=E_{k,0}(0)-E_F(0)+2kh_1,
\end{equation}
and positivity of $\delta_k$ is seen from Fig.~\ref{fig:z2-easy-plane}(c). Here $E_F(0)=0$ in our normal-ordering convention; retaining it explicitly makes the subtraction independent of an additive energy offset.

\section{Pinned Goldstone comparison}
\label{sec:Goldstone}

In the broken phase, $\Phi=(v+\sigma)e^{i\theta}$. Eliminating the massive radial mode gives:
\begin{equation}\label{eq:phase-action}
\cL_\theta=\frac{K_t}{2}(\partial_\tau\theta)^2+\frac{K_s}{2}(\boldsymbol\nabla\theta)^2+\cdots.
\end{equation}
Adding a pinning term of order $R^{-2}$ gives the pseudo-Goldstone with the conformal mass. The canonically normalized phase fluctuation is $\phi=(K_t K_s)^{1/4}(\theta-\theta_0)$, where $\theta_0$ is the pinned orientation. We will compare our model to He's model \cite{He2025FreeScalar}. In the conventions of the Letter:
\begin{align}
\label{eq:He-H}H_{\He}&=\int\dd\Omega\left[(n^0)^2+U n^z\boldsymbol\nabla^2n^z+\frac{h_x}{R^2}n^x\right],\\
\label{eq:EP-H}H_{\ep}&=\int\dd\Omega\left[(n^0)^2-U(n^x\boldsymbol\nabla^2n^x+n^y\boldsymbol\nabla^2n^y)+\frac{h_x}{R^2}n^x\right].
\end{align}
Their difference is the $SU(2)_F$-invariant derivative operator:
\begin{equation}\label{eq:EP-He-difference}
H_{\ep}-H_{\He}=-U\int\dd\Omega\sum_{a=x,y,z}n^a\boldsymbol\nabla^2n^a.
\end{equation}
The energy units are fixed by $\Delta_\phi=1/2$, so the shifted and rescaled spectrum is:
\begin{equation}
d_a^X(N)=\frac{E_a^X(N)-E_0^X(N)}{2[E_\phi^X(N) - E_0^X(N)]},\quad X\in\{{\rm ep}, {\rm He}\}.
\end{equation}
The calibration levels are $\mathcal{C}_6= (\phi^2, \partial\phi, \partial^2\phi, \partial\phi^2, \partial^2\phi^2, T_{\mu\nu})$, whose exact conformal dimensions are $\boldsymbol{\Delta}=(1, \frac32, \frac52, 2, 3, 3)$, respectively. The optimization over $(U, h_x)$ is done by minimizing the root mean square (RMS) cost function:
\begin{equation}
\mathcal{Q}_X(N) = \sqrt{\frac16 \sum_{a\in\mathcal{C}_6}[d_a^X(N) - \Delta_a]^2}.
\end{equation}
We then compare the $H_\ep$ and $H_\He$ spectra beyond the calibration levels by picking a set $\mathcal{S}_{15}$ of the next fifteen lowest levels (outside the calibration set), and computing there RMS:
\begin{equation}
{\rm RMS}(N)=\sqrt{\frac1{15}\sum_{a\in\mathcal{S}_{15}}[d_a^\ep(N) - d_a^\He(N)]^2}.
\end{equation}
The $H_\ep$ parameters were optimized at $N_{\rm ref}=12$,
\begin{equation}
(U,h_x)_{\ep}=(0.1868309361,0.3707467799),
\qquad
(U,h_x)_{\He}=(0.18,0.42),
\label{eq:Goldstone-params}
\end{equation}
and then held fixed for $N=8,10,12,14,16$. To reliably identify the levels, we label them by:
\begin{equation}\label{eq:level-labels}
(\PH,\mathbb{Z}_2,R_y,L,\text{sector level}).
\end{equation}
The first four entries are quantum numbers labeling the symmetry sectors, and within each sector, the ``sector level'' labels states in the increasing energy order (up to level crossing). The calibration states are summarized in Table~\ref{tab:Goldstone-states}.
\begin{table}[h]
\centering
\begin{tabular}{lcccccc}
	\toprule
	state & $\PH$ & $\mathbb{Z}_2$ & $R_y$ & $L$ & sector level & dimension\\
	\midrule
	$\phi$ & $+$ & $-$ & $+$ & 0 & 1 & 1/2 \\
	$\partial\phi$ & $+$ & $-$ & $-$ & 1 & 1 & 3/2 \\
	$\partial^2\phi$ & $+$ & $-$ & $+$ & 2 & 1 & 5/2 \\
	$\phi^2$ & $+$ & $+$ & $+$ & 0 & 2 & 1 \\
	$\partial\phi^2$ & $+$ & $+$ & $-$ & 1 & 1 & 2 \\
	$\partial^2\phi^2$ & $+$ & $+$ & $+$ & 2 & 1 & 3 \\
	$T_{\mu\nu}$ & $+$ & $+$ & $+$ & 2 & 2 & 3 \\
	\bottomrule
\end{tabular}
\caption{The quantum numbers help identify calibration states as these lowest states in the respective symmetry sectors.}
\label{tab:Goldstone-states}
\end{table}

We then pick the common levels selected by $\max_{N\in\{8,10,12,14,16\}}\frac{d_a^\ep(N) + d_a^\He(N)}{2} \leq 3.5$, which is how we get the 15-level set $\mathcal{S}_{15}$. The energy window $3.5$ is chosen empirically as low enough to probe the IR physics, where the models $H_\ep$ and $H_\He$ are expected to agree (the agreement worsens for wider energy windows due to the irrelevant interaction \eqref{eq:EP-He-difference}). The numerical results show that both the calibration costs $\mathcal{Q}_X$ and the 15-level RMS decrease monotonically:
\begin{table}[h]
\centering
\begin{tabular}{c@{\qquad}cc@{\qquad}c}
	\toprule
	$N$ & $\mathcal Q_{\He}$ & $\mathcal Q_{\ep}$ & common 15-level RMS \\
	\midrule
	8  & 0.17301 & 0.12435 & 0.11898 \\
	10 & 0.14518 & 0.09454 & 0.11004 \\
	12 & 0.12483 & 0.07731 & 0.10311 \\
	14 & 0.10931 & 0.06857 & 0.09803 \\
	16 & 0.09712 & 0.06549 & 0.09467 \\
	\bottomrule
\end{tabular}
\caption{Free scalar calibration cost and models' difference. The final column uses the same 15 noncalibration states at every size, selected by quantum numbers and mean normalized energy below $3.5$.}
\label{tab:Goldstone-results}
\end{table}\\
The two calibration cost functions are shown in Fig.~\ref{fig:goldstone-comparison}(a) of the Letter's End Matter, and the 15-level RMS plot is in Fig.~\ref{fig:goldstone-comparison}(b). Finally, the broader $N=16$ comparison data in Fig.~\ref{fig:goldstone-comparison}(c) displays 41 noncalibration states (along with the 7 calibration states) with both energies below $4.5$, depicting $d_a^\ep(16)$ vs $d_a^\He(16)$, labeled as $\Delta_\ep$ and $\Delta_\He$ there. That the states cluster along the diagonal again demonstrates that $H_\ep$ and $H_\He$ are very close.

The eigensolver tolerance was $10^{-9}$, with degeneracy tolerance $2\times10^{-9}$ and $L^2$ tolerance $5\times10^{-4}$. Numerical diagonalization errors are far below the plotted symbols. The meaningful uncertainties are systematic, such as calibration states choice, finite-size corrections, and possible branch exchanges among states with identical quantum numbers. Interpolating the two Hamiltonians and tracking states overlaps would strengthen the claim.

\end{document}